\documentclass[12pt]{article}
\usepackage[left=1in,top=1in,right=1in,bottom=1in,nohead]{geometry}

\usepackage{url,subfigure,latexsym,amsmath,amssymb, amsfonts,enumitem,
  epsfig, graphics,times, amsthm,mathtools, bbm}

\graphicspath{{./Images/}}
\usepackage[normalem]{ulem}

\usepackage{tikz, pgfplots}
\usetikzlibrary{automata,positioning,arrows, intersections, calc}
\pgfplotsset{compat=1.8}

\usepackage{cite}
\usepackage{wrapfig}
\usepackage{xcolor}
\usepackage{nameref}
\usepackage{listings}
\usepackage[colorlinks]{hyperref}
\usepackage{makeidx}
\usepackage{tabularx}
\usepackage{array}
    \newcolumntype{P}[1]{>{\centering\arraybackslash}p{#1}}
    \newcolumntype{M}[1]{>{\centering\arraybackslash}m{#1}}

    \providecommand{\keywords}[1]{\textbf{Keywords:} #1}
     \providecommand{\classification}[1]{\textbf{AMS subject classifications:} #1}

\newtheorem{algorithm}{Algorithm}

\newtheorem*{thm*}{Theorem}

\theoremstyle{definition}
\newtheorem{definition}{Definition}[section]

\newtheorem{example}{Example}[section]
\newtheorem{remark}{Remark}

\newcommand{\E}{\mathbb E}

\newcommand{\I}{\mathcal I}

\newcommand{\RR}{\mathbb R}

\newcommand{\ZZ}{\mathbb Z}

\newcommand{\Sp}{\mathcal S}
\newcommand{\C}{\mathcal C}
\newcommand{\Reac}{\mathcal R}

\newcommand{\K}{\mathcal K}

\newcommand{\Fsim}{F^{\mathrm{sim}}}
\newcommand{\Fsimtheta}{F^{\mathrm{sim},\theta}}
\newcommand{\Fsimthetah}{F^{\mathrm{sim},\theta+h}}

\title{Parametric Sensitivity Analysis for Models of Reaction Networks within Interacting Compartments}

\author{
David F. Anderson\thanks{Department of Mathematics, University of
  Wisconsin-Madison, USA.  anderson@math.wisc.edu, grant support from NSF-DMS-2051498. Corresponding author. },
\and 
Aidan S. Howells\thanks{Dipartimento di Scienze Matematiche, Politecnico di Torino, Turin, Italy. aidan.howells@polito.it,
support from MUR PRIN grant number 2022XRWY7W.}
}

\begin{document}

\maketitle

\begin{abstract}
Models of reaction networks within interacting compartments (RNIC) are a generalization of stochastic reaction networks.   It is most natural to think of the interacting compartments as ``cells''  that can appear, degrade, split, and even merge, with each cell containing an evolving copy of the underlying stochastic reaction network.  Such models have a number of parameters, including those associated with the internal chemical model and those associated with the compartment interactions, and it is natural to want efficient computational methods for the numerical estimation of sensitivities of model statistics with respect to these parameters.  Motivated by the extensive work on computational methods for parametric sensitivity analysis in the context of stochastic reaction networks over the past few decades,  we provide a number of methods in the basic RNIC setting.  Provided methods include the (unbiased) Girsanov transformation method (also called the Likelihood Ratio method) and a number of coupling methods for the implementation of finite differences, each motivated by methods from previous work related to stochastic reaction networks.  We provide several numerical examples comparing the various methods in the new setting.  We find that the relative performance of each method is in line with its analog in the ``standard'' stochastic reaction network setting.  
We have  made all of the Matlab code used to implement the various methods freely available for download.
\end{abstract}


\classification{60J27, 60J28, 60H35, 65C05}

\section{Introduction}
The last few decades have seen a large amount of research focused on utilizing stochastic reaction networks to understand myriad processes, including gene regulatory networks, viral and bacterial infections, and more \cite{Yin2002, Arkin1998, Becskei2005, Elowitz2002, Paulsson2011,Maamar2007, Paulsson2004, Raser2004,ULPOSP2016}.  The standard mathematical model for a stochastic reaction network treats the system as a  continuous-time Markov chain on $\ZZ^d_{ \ge 0}$, where $d$ is the number of species of the system, with reactions determining the state transitions of the model.  Such mathematical models are typically used under the assumption that the real-world system being studied resides within a ``well-stirred'' environment.
See \cite{AndKurtz2011, AK2015} for a general (mathematical) introduction to such models.  These models are often simulated via the so-called ``Gillespie algorithm'' named after Dan Gillespie who wrote some of the seminal papers on these models in the mid-1970s \cite{Gill76,Gill77}.

The assumption that the system of interest resides within a well-stirred environment can be generalized in numerous ways, depending on the problem being studied. For example, in some situations it may be natural to track the position of individual molecules, either in a discretized space or a continuous space setting \cite{doi1976stochastic,agbanusi2014comparison, Sam2013, erban2014special}.  
Another approach, and the one we take here,  is to assume the existence of multiple interacting ``compartments'' (or cells) each of which contains an evolving copy of a given stochastic reaction network.  This approach  is useful in numerous modeling scenarios including  
the dynamics of membrane-bound organelles \cite{vagne2018stochastic},  the study of clustering proteins \cite{saunders2015aggregation}, 
and the dynamics of clonal cells during their development \cite{rulands2018universality}.
In \cite{DZ}, a  modeling framework was formalized for this type of system which can be briefly summarized in the fallowing manner.
\begin{itemize}
\item There are a number of ``compartments'' (one can think of them as cells) that are interacting dynamically.  Following the language of \cite{DZ}, the four types of interactions for the compartments are: (i) inflows (compartments spontaneously appear at some rate), (ii) outflows (compartments spontaneously disappear, or die), (iii) coagulation or the merger of two compartments, and (iv)  fragmentation (a compartment divides into two compartments).

\item Each compartment contains an evolving copy of a given stochastic reaction network.  The stochastic reaction networks evolve independently between compartment events/interactions.

\item When a compartment appears, the initial state of its contents is chosen from a given distribution on $\ZZ^d_{\ge 0}$.  We will later denote this distribution by $\mu$.
\item When a compartment disappears, its contents disappear as well.
\item When compartments merge, their contents combine.
\item When a compartment fragments, the contents are randomly split between the two compartments.
\end{itemize}
A first comprehensive mathematical analysis of these models (with results related to such basic questions as transience, recurrence, explosiveness, stationary behavior, etc.) can be found in \cite{anderson2023stochastic}.  We note that a different model, also considered in \cite{DZ}, allows for the rates associated with compartment events (such as mergers and fragmentations) to depend upon the contents of the different compartments.  Such models are not considered within this paper.  Extending the methods presented here to that modeling scenario is a reasonable avenue of future work, as pointed out in our Discussion Section \ref{sec:discussion}.

These RNIC models have many parameters: those associated with the reaction network (evolving within each compartment), the parameters associated with the evolving compartment model, the parameters of the initial distribution for when compartments appear, etc.  When working with any mathematical model a critical question is: how sensitive is the model to perturbations in its parameters?  As we are dealing with a stochastic process, the most natural way to pose this question is as follows: what are the derivatives of the relevant expectations of the model?  Here the ``relevant expectations'' will depend upon the model being studied and the questions being posed, but could include expected numbers of certain species (summed over all the compartments, perhaps) at a given time, the probability of the model being in a certain region of state space at a given time, etc.  The study of such derivatives is typically called \textit{parametric  sensitivity analysis} and plays a critical role in the computational study of stochastic models in all fields that utilize them \cite{GlynnAsmussen2007, Komorowski2011, Plyasunov2007, AndCFD2012, mclure2017healthcare, amilo2024modeling, ruess2017sensitivity}.  

In this paper, we provide a number of computational methods that estimate derivatives of expectations of RNIC models.  In particular, we provide a Girsanov transformation method (also termed a likelihood transformation method), which is unbiased, and a number of finite difference methods. Finite difference methods require a perturbation parameter $h$, where smaller values of $h$ reduce bias but increase variance (see Appendix \ref{sec:MonteCarlo}).  We do not provide any pathwise differentiation methods (also termed \textit{Infinitesimal Perturbation Analysis}) as stochastic reaction networks (and, hence, RNIC models) nearly always have ``interruptions,'' ensuring these methods are not valid in this context \cite{glasserman1990gradient, wolf2015hybrid}. 
The finite difference methods provided here are each based off a different coupling strategy taken from the computational stochastic reaction network literature.  
These couplings are: Common Random Variables (CRV)
(i.e., using the same seeds for the random number generator), the Common Reaction Path (CRP) method \cite{Khammash2010}, the local Common Reaction Path method (local-CRP)\cite{anderson2015asymptotic}, and the Split Coupling method \cite{AndCFD2012, AndHigham2012}.   We do not attempt to extend other methods from the stochastic reaction network literature, such as the Auxiliary Path Algorithm \cite{gupta2013unbiased}, the Poisson Path Algorithm \cite{gupta2014efficient}, the hybrid pathwise approach\cite{wolf2015hybrid} or the  Integral Path Algorithm \cite{gupta2018estimation} as they are beyond the scope of this article (which mostly focuses on finite difference methods).  However, these can sometimes be low variance, unbiased alternatives to the Girsanov transformation method  and extending them would be a reasonable project in the future, as we mention in our Discussion Section \ref{sec:discussion}.

We provide a number of numerical examples, with perturbations in both the chemical and compartment parameters.  
Our general conclusions follow similar conclusions from parametric sensitivity analysis for ``standard'' stochastic reaction networks.  Specifically, because the Girsanov method is unbiased and easy to implement, it should be attempted first.  However, if the Girsanov method provides variances the are too large for a particular computational budget,  then a finite difference method should be implemented.  In that case, the Split Coupling is a good method to choose when one parameter, or a small number of parameters, is being perturbed.  However, if a large number of parameters are being perturbed, then the CRP methods could be the most efficient. 


The remainder of the paper is arranged as follows.    In Section \ref{sec:models}, we formally introduce the relevant mathematical models, including both the standard model of stochastic reaction networks and the RNIC model.  In Section \ref{sec:methods}, we provide various computational methods for the estimation of parametric sensitivities.  In particular, we demonstrate how to implement the Girsanov transformation method and the Split Coupling in the RNIC context.  
 In Section \ref{sec:examples}, we provide a number of computational examples.  Finally, in Section \ref{sec:discussion}, we discuss our findings and suggest some avenues of future work.
 In the appendix Section \ref{sec:MonteCarlo}, and for the sake of completeness, we give a very brief introduction to Monte Carlo methods in the context of estimating parametric sensitivities with finite difference methods.  In the appendix Section \ref{sec:MC_Algorithms}, we provide a number of algorithms that are utilized for the standard stochastic reaction network models (as introduced in Section \ref{sec:models}).

\section{Mathematical models}
\label{sec:models}

We first introduce the standard model for stochastic reaction networks in Section \ref{sec:stoch_reac_network}.  Next, in Section \ref{sec:RNIC}, we introduce the RNIC model.  We also provide in Algorithm \ref{alg:exact} a pseudo-algorithm for generating a single trajectory of an RNIC model.

\subsection{Stochastic reaction networks}
\label{sec:stoch_reac_network}

We begin with the definition of a reaction network.

\begin{definition}
A \textit{reaction network} is a triple of non-empty, finite sets, usually denoted $\{\Sp,\C,\Reac\}$.
\begin{enumerate}[itemsep=-.05ex]
\item \textit{Species}, $\Sp$: the  components whose abundances we wish to model dynamically.
\item \textit{Complexes}, $\C$: non-negative integer linear combinations   of the species.
\item \textit{Reactions}, $\Reac$:   a binary relation on the complexes.  The relation is often denoted   ``$\to$''.
\end{enumerate} 
For $y,y'\in \C$ with $y\to y' \in \Reac$, we call $y$ and $y'$ the \textit{source} and \textit{product} complexes of that reaction, respectively. 
 We write $\emptyset$ for the complex with all zero coefficients.\hfill $\triangle$
\end{definition}

An example  demonstrates the terminology. 
\begin{example}\label{example:8889901}
Consider a standard model of transcription and translation, with protein dimerization (see Example 2.4 in  \cite{AK2015}):  
\begin{align}
\emptyset & \to M \tag{Transcription}\\
M & \to M +P \tag{Translation}\\
M & \to \emptyset \tag{Degradation of mRNA}\\
P & \to \emptyset\tag{Degradation of protein}\\
2P& \to D \tag{Dimerization}\\
D & \to \emptyset,\tag{Degradation of dimer}
\end{align}
where $M$ represents mRNA, $P$ represents protein, and $D$ represents dimer.  We are assuming the cell has one gene, and so we suppress that species.  

For this model, the set of species is $\mathcal{S} = \{M,P,D\},$ the set of complexes is $\C = \{\emptyset, M, P, M+P, 2P, D \},$
and the six reaction types are as above, $\Reac = \{\emptyset \to M, \ M \to M+P,\ M\to \emptyset, \dots\}$. \hfill $\square$
\end{example}

Returning to the general case, we fix notation and assume there are $d$ species, which we label as  $\{S_1,\dots, S_d\}$. 
We will denote the process by $X$, so that $X(t) \in \ZZ^d_{\ge 0}$ gives the state at time $t \ge 0$.  Therefore, $X(t)$ is a vector  giving the abundance of
molecules of each species at the specified time.  The state transitions for the model are then determined by the different reactions.  For the $k$th reaction, we let $y_k\in \ZZ^d_{\ge0}$ and $y_k'\in \ZZ^d_{\ge0}$ be the vectors whose $i$th component gives the multiplicity of species $i$ in the source and product complexes, respectively, and let $\lambda_k:\ZZ^d_{\ge 0} \to \RR_{\ge 0}$  give the \emph{transition intensity}, or rate,
at which the reaction occurs. (Note that in the biological and chemical literature
transition intensities are referred to as \textit{propensities} \cite{Gill76,Gill77}.)
If the $k$th reaction occurs at time $t$, then the old state, $X(t-)$, is  updated
by addition of the \textit{reaction vector} 
\begin{align*}
\zeta_k := y_k' - y_k
\end{align*}
 and the new state is given as $X(t) = X(t-) + \zeta_k$.  Here, by $X(t-)$ we mean the left-sided limit $\lim_{s\to t^-}X(s)$.
  For example, if a model consists of the three species $\{S_1,S_2,S_3\}$ and if  the
reaction type $S_1 + S_2 \to S_3$ takes place, then we would update with  
\[
y_k = \left[\begin{matrix}1\\1\\0 \end{matrix}\right], \quad y_k' = \left[\begin{matrix}0\\0\\1\end{matrix}\right], \quad \text{ and } \quad \zeta_k =  \left[\begin{array}{r} 
		-1\\
		-1\\
		1
		\end{array}\right].
\]

Denoting the number of times  the $k$th reaction occurs by time $t$ as $R_k(t)$, simple bookkeeping implies 
\[
    X(t)= X(0) + \sum_k R_k(t) \zeta_k,
\]
where the sum is over  reactions. Note that
each $R_k$ is a counting process (starts at 0 and can only change by increases of size $+1$) with intensity  $\lambda_k(X(t))$.   Letting $\{Y_k\}$ be independent unit-rate Poisson processes, Kurtz's representation (which is useful for both analysis and computation) has $R_k(t) = Y_k(\int_0^t \lambda_k(X(s)) ds)$ \cite{Kurtz78,AK2015}.  Hence, the model satisfies the  following equation, which is often termed the \textit{random time change representation} of Kurtz:
\begin{align}\label{eq:RTC}
X(t) = X(0) + \sum_{k} Y_k \left( \int_0^t \lambda_k(X(s)) ds\right) \zeta_k,
\end{align}
where the sum is over the reaction types.
We will denote a family of such models, parameterized by $\theta$ (which, in the case of stochastic mass-action kinetics---see below in \eqref{eq:stochasticMA}---is most likely a rate constant, $\kappa_k$, of the system), as 
\begin{equation}\label{eq:parameterized_RTC}
X^\theta(t) = X^\theta(0) + \sum_k Y_k \left( \int_0^t \lambda_k^\theta (X^\theta(s)) ds \right) \zeta_k.
\end{equation}

The most common, though certainly not the only, choice of intensity function $\lambda_k$ is that of stochastic \textit{mass-action kinetics}. The stochastic form of the law of mass-action says that for some
constant $\kappa_k$ the rate of the $k$th reaction is
\begin{equation}\label{eq:stochasticMA}
	\lambda_k(x) = \kappa_k \prod_{i} \frac{x_{i}!}{(x_{i}  - y_{k,i})!}
\end{equation}
where $y_{k,i}$ is the $i$th component of the source complex $y_k$.\footnote{We note that in \cite{anderson2023stochastic}, the intensity functions were defined slightly differently as $\lambda_k(x) = \kappa_k \prod_{i} \frac{x_i}{y_{k,i}!(x_i-y_{k,i})!}.$} 
The rate constants are typically placed next to their  reaction arrow in the reaction graph, as in \eqref{eq:compartment8907} or \eqref{eq:867986}, which are found later in the paper.

\subsection{Reaction Network within Interacting Compartments (RNIC)}
\label{sec:RNIC}

We have already introduced the basic idea of the RNIC model in the introduction.  Here we  fill in the necessary technical details.  We point the interested reader to \cite{anderson2023stochastic} for a full mathematical introduction  for the model.  Note that the detailed choices we specify in this section, and especially in the pseudo-code provided at the end, are those that we make in our Matlab code for the simulation of these processes (and which is freely available).

As mentioned in the introduction, the RNIC model consists of interacting compartments, each of which contains an evolving copy of a given reaction network.  Therefore, we begin by denoting the reaction network by 
\[
\I = \{\Sp, \C, \Reac\}.
\]
When assuming stochastic mass-action kinetics, we will denote the set of rate constants by $\K$, and denote 
\[
\I_{\K} = \{\Sp, \C, \Reac, \K\}.
\]

Turning to the compartment model, the number of compartments is modeled as a stochastic reaction network as well.  Specifically, by
\begin{equation}\label{eq:compartment8907}
\emptyset \overset{\kappa_I}{\underset{\kappa_E}{\rightleftarrows}} C \overset{\kappa_F}{\underset{\kappa_C}{\rightleftarrows}} 2C,
\end{equation}
with stochastic mass-action kinetics and where the rate constants have been placed next to the respective reaction arrows with $\kappa_I$ the rate constant for inflows, $\kappa_E$ for exits, $\kappa_F$ for fragmentations, and $\kappa_C$ for coagulations (following the terminology of \cite{DZ} and \cite{anderson2023stochastic}).  We will denote by 
$\mathcal H_{\K}$ the compartment reaction network \eqref{eq:compartment8907} (where we have also expanded the set $\K$ to include the rate constants for both the reaction network $\I$ and the compartment model).

Denote by $M_C(t)$ the number of compartments at time $t\ge 0$.  Then, following \eqref{eq:RTC},  the compartment model satisfies the  stochastic equation
\begin{align*}
    M_C(t) &= M_C(0) + Y_I\left( \kappa_I t\right) - Y_E\left( \kappa_E \int_0^t M_C(s) ds\right) + Y_F\left( \kappa_F \int_0^t M_C(s) ds\right)\\
    &\hspace{.2in} - Y_C\left( \kappa_C \int_0^t M_C(s) (M_C(s) - 1) ds\right),
\end{align*}
where $Y_I, Y_E, Y_F,$ and $Y_C$ are independent unit-rate Poisson processes.

Between the transition times of the Markov process $M_C$ each compartment contains an independently evolving copy of the stochastic reaction network $\mathcal I_{\K}$.  We numerically order the compartments and for $i \in \{1, \dots, M_C(t)\}$ we denote the state of the stochastic reaction network evolving in the $i$th compartment by $X_i(t)$.
We now simply need to specify what happens to the model at the transition times of $M_C$.  We do that below.  We point out  that all choices of random variables below are independent from each other and independent from the Poisson processes $Y_I, Y_E, Y_F$, and $Y_C$ mentioned above.

We will assume that a transition for $M_C$ takes place at time $t$.  Hence, before the transition there are exactly $M_C(t-)$ compartments (where, again, $t-$ denotes a limit in $t$ from the left).  We now consider each of the four possible transition types for $M_C$. We note that the names of certain indices below, including IndexDel, IndexFrag, ri1, and ri2, are the names of the indices utilized in the Matlab code we are making available.
\begin{enumerate}
    \item Case 1:  a new compartment arrives at time $t$. In short, in this case we append a new compartment at the end of our list and initialize the stochastic reaction network via a probability distribution we denote by $\mu$. Specifically, we do the following.
    \begin{enumerate}
        \item Set $M_C(t) = M_C(t-) + 1$.
        
        \item Set $X_i(t) = X_i(t-)$ for all $i \in \{1, \dots, M_C(t-)\}$ (that is, the reaction networks within each of the existing compartments remain the same).
        
        \item Initialize the stochastic reaction network of the new compartment. The state of the new compartment will be drawn from a fixed probability measure $\mu$ on $\ZZ^d_{\ge 0}$, where $d$ is the number of species in the reaction network model $\I_\K$.
    \end{enumerate}
    \item Case 2: a compartment exits at time $t$.  In short, in this case, we delete a compartment at random and shift the indices of the remaining compartments down appropriately.  Specifically, we do the following.
    \begin{enumerate}
        \item Choose from $\{1,\dots, M_C(t-)\}$ uniformly;  with the chosen index called IndexDel.
        \item For $i \in \{1, \dots, \text{IndexDel}-1\}$,  set $X_i(t) = X_i(t-)$.
        \item For $i \in \{\text{IndexDel},\dots, M_C(t-)-1\}$,  set $X_i(t) = X_{i+1}(t-)$ (that is, we shift indices by 1).
        \item Set $M_C(t) = M_C(t)-1$.
    \end{enumerate}
     
    \item Case 3: a compartment fragments into two at time $t$.  In short, in this case, one compartment is chosen to divide.  Each molecule of each species is chosen to either remain in the old compartment, or be placed into a new compartment numbered as $M_C(t-) + 1$.  Specifically, we do the following.
    \begin{enumerate}
        \item Set $M_C(t) = M_C(t-)+1$.
        \item Choose from $\{1,\dots, M_C(t-)\}$ uniformly, with the chosen index called IndexFrag.
        \item For each species $S_j \in \Sp$, we generate a binomial random variable with parameters $n = X_{\text{IndexFrag},j}(t-)$ (abundance of species $S_j$ in compartment IndexFrag) and $p = 1/2$.  
        \item The values given by the binomial random variables determine the state of $X_{\text{IndexFrag}}(t)$ (i.e., they are the molecules that remain).
        \item The state of $X_{M_C(t)}(t)$ is $X_{\text{IndexFrag}}(t-) - X_{\text{IndexFrag}}(t).$
    \end{enumerate}
    
    \item Case 4: a coagulation (merger). In short, in this case two compartments are chosen to merge.  The contents are combined into one of the compartments and the other is deleted.  Specifically, we do the following.
    \begin{enumerate}
        \item Choose two indices, $\text{ri1}, \text{ri2}\in \{1,\dots, M_C(t-)\}$ uniformly, with $\text{ri1}\ne \text{ri2}$.
        \item Combine the contents into compartment $\text{ri1}$: $X_{\text{ri1}}(t) = X_{\text{ri1}}(t-) +X_{\text{ri2}}(t-).$
        \item Delete the compartment $\text{ri2}$ as detailed in Case 2 above.
        \item Set $M_C(t) = M_C(t-)-1$.
    \end{enumerate}
\end{enumerate}

In agreement with \cite{anderson2023stochastic}, we will denote the stochastic process associated to the full RNIC model by $\Fsim$ (where the label ``sim'' denotes that this representation of the model is natural for simulations).  Hence, $\Fsim(t)$ determines $M_C(t)$ and each of $X_i(t)$, $i \in \{1,\dots, M_C(t)\}$, for each $t \ge 0$. The process $\Fsim$ is a continuous-time Markov chain with discrete state space $\bigcup_{m=0}^\infty\left(\ZZ^d_{\ge 0}\right)^m$, the space of finite tuples of $\ZZ^d_{\ge 0}$ (Lemma 2.5 in \cite{anderson2023stochastic}).  For example, if $d=2$ and $M_C(t)= 3$ a possible realization of $\Fsim(t)$ is
\[
\Fsim(t) = \left( \left[\begin{array}{c}
4\\
5
\end{array}
\right], \left[\begin{array}{c}
0\\
10
\end{array}
\right], \left[\begin{array}{c}
17\\
2
\end{array}
\right]
\right)
\]
which would have $X_1(t) = \left[\begin{array}{c}
4\\
5
\end{array}
\right],$ $X_2(t) = \left[\begin{array}{c}
0\\
10
\end{array}
\right]$, and $X_3(t) = \left[\begin{array}{c}
17\\
2
\end{array}
\right]$.  When we wish to denote a parameterized family, we will once again append our processes with a $\theta$: $\Fsimtheta$, $M_C^\theta$, and $X_i^\theta$.

Having specified the model, we can present a pseudo-code for how to simulate a given RNIC process.  In the pseudo-code below, we do not specify the algorithms being used to simulate $M_C$ or the  stochastic reaction networks within each compartment, as we leave that up to the implementer.  Natural choices are the Gillespie algorithm \cite{Gill76,Gill77} or the next reaction method \cite{Anderson2007a,Gibson2000}.  These can both be found in Appendix \ref{sec:MC_Algorithms}. We will denote the initial distribution of $M_C$ by $\nu$, the initial distribution for the initial compartments by $\mu_0$, and the initial distribution for compartments that arrive as $\mu$. Note that it is possible, though not necessary, that $\mu_0=\mu$.

\begin{algorithm}[Pseudo-code for simulating an RNIC model, $\Fsim$]
\label{alg:exact}
Given: a stochastic reaction network $\I_{\K}$, the compartment model \eqref{eq:compartment8907}, the parameters of the model $\K$, which consist of the rate constants $\{\kappa_k\}$ for $\I_{\K}$, the parameters of the compartment model $\{\kappa_I, \kappa_E, \kappa_F, \kappa_C\}$, and initial distributions $\nu$, $\mu_0$, and $\mu$.

Repeat steps 4 -- 8 until a stopping criteria is reached.  All calls to random variables are independent from all others.
\begin{enumerate}[itemsep=-.05ex]
    \item Determine $M_C(0)$ via $\nu$.
    \item For each $i \in \{1, \dots, M_C(0)\}$ determine $X_i(0)$ via $\mu_0$.  
    \item Set $t = 0$.
    \item Using an exact simulation method (e.g., Gillespie's algorithm---Algorithm \ref{alg:Gillespie}---or the next reaction method---Algorithm \ref{alg:NRM}), determine the time $\Delta$ until the next transition of the compartment model, $M_C$.
    \item Using an exact simulation method (e.g., Gillespie's algorithm---Algorithm \ref{alg:Gillespie}---or the next reaction method---Algorithm \ref{alg:NRM}), for each $i \in \{1, \dots, M_C(t)\}$ simulate $X_i$ from time $t$ to time $t + \Delta$.
    \item Determine which type of transition occurs for the compartment model at time $t + \Delta$.
    \item Update the model as detailed in the four cases listed above depending upon which type of transition takes place for the compartment model.
    \item Set $t \leftarrow t + \Delta$.
\end{enumerate}
\end{algorithm}

\section{Methods for parametric sensitivities}
\label{sec:methods}

We provide a number of methods for the estimation of parametric sensitivities of RNIC models.  In Section \ref{sec:girsanov}, we provide the (unbiased) Girsanov or Likelihood Ratio method in this setting.  In Section \ref{sec:FD}, we provide a number of coupling methods for the implementation of finite differences.

\subsection{Likelihood Ratio, or Girsanov, transformation method}
\label{sec:girsanov}

We  provide the Likelihood Ratio method \cite{GlynnAsmussen2007}, often called the Girsanov transformation method in the biosciences \cite{Plyasunov2007}, for the RNIC model when the perturbed parameters are rate constants for either the compartment model, $\mathcal{H}_\K$, or the chemical model $\I_\K$.  There are other possible parameters to consider, such as any parameters associated with the  distributions $\nu$, $\mu_0$, or $\mu$.  However, those cases can be handled in a similar, and straightforward, manner.  

While we do not provide formal mathematical justification for this method here (though we refer the interested reader to Refs. \cite{wang2021validity, GlynnAsmussen2007, warren2012steady}) we do try to provide some intuition by explaining how the method works for the estimation $\frac{d}{d\theta} \E[f(X^\theta(t))]$, where $X^\theta$ is a  standard stochastic reaction network and $t$ is some  terminal time.  Extension to the RNIC model is then relatively straightforward.

Consider a reaction network with intensity functions $\{\lambda_k^\theta\}$, which are parameterized by $\theta$, and associated jump directions $\{\zeta_k\}$.  Denote $\lambda_{\text{tot}}^\theta(x) = \sum_k \lambda_k^\theta(x)$. The key point for this method is that it proceeds by finding an appropriate function $W^\theta$, often called the ``weight function,'' so that 
\[
\frac{d}{d\theta} \E[ f(X^\theta(t))] = \E[f(X^\theta(t))W^\theta(t)].
\]
One can then estimate the derivative via Monte Carlo with  independent samples of $f(X^\theta(t))W^\theta(t)$.

The function $W^\theta$ is found by taking the derivative, with respect to the parameter $\theta$, of the logarithm of the density of the process up to time $t$.  For example, suppose that $X:[0,t]\to \ZZ^d_{\ge 0}$ is a particular path of the process (that one may have simulated, for example).  Then, for this particular path,  let $t_0 = 0$, $N(t)$ be the number of reactions that have taken place up to time $t$, $t_i$ be the time of the $i$th transition (for $i \ge 1$),  $k_i^*$ be the index of the reaction type that takes place at time $t_i$, and $\Delta_{i} :=t_{i+1}-t_i$ be the holding time in state $X_{t_i}$.
Thinking in terms of the Gillespie algorithm, i.e., which reaction happens next and when that reaction takes place, which are a discrete event and an exponential holding time, respectively, it is then relatively straightforward to see \cite{GlynnAsmussen2007, Plyasunov2007} that the density of the process at time $t\ge 0$  is proportional to
\begin{align*}
\Bigg(\prod_{i = 1}^{N(t)}\underbrace{\frac{\lambda^\theta_{k_i^*}(X(t_{i-1})}{\lambda_{\text{tot}}^\theta(X(t_{i-1}))}}_{\text{Which reaction}} & \underbrace{\lambda_{\text{tot}}^\theta(X(t_{i-1})) e^{-\lambda_{\text{tot}}^\theta(X(t_{i-1}))\cdot \Delta_{{i-1}}}}_{\text{Exponential holding time}}\Bigg) \cdot \underbrace{e^{-\lambda_{\text{tot}}^\theta(X(t_{N(t)}))(t-t_{N(t)})}}_{\text{No reaction from $t_{N(t)}$ to $t$}}\\
&= \left( \prod_{i = 1}^{N(t)} \lambda^\theta_{k_i^*}(X(t_{i-1})) \right)e^{-\int_0^t \lambda_{\text{tot}}^\theta(X(s)) ds}.
\end{align*}
Taking logarithms and derivatives is now straightforward, and we have
\begin{align*}
    W^\theta(t) = \sum_{i = 1}^{N(t)} \frac{\frac{d}{d\theta} \lambda^\theta_{k_i^*}(X(t_{i-1})) }{\lambda^\theta_{k_i^*}(X(t_{i-1}))} - \int_0^t \frac{d}{d\theta} \lambda_{\text{tot}}^\theta( X(s)) ds.
\end{align*}
Note that in the case of stochastic mass-action kinetics with $\theta = \kappa_j$, we have the simple expressions
\begin{align*}
    \frac{\frac{d}{d\theta} \lambda^\theta_{k_i^*}(x) }{\lambda^\theta_{k_i^*}(x)} = \begin{cases}
        \frac{1}{\kappa_j} & \text{ if }  k_i^*= \kappa_j\\
        0 & \text{ else}
    \end{cases},\quad \text{and} \quad \frac{d}{d\theta} \lambda_{\text{tot}}^\theta(x) = \frac{1}{\kappa_j} \lambda_j^\theta(x).
\end{align*} 
We point out that in the case of mass-action kinetics the above terms are not defined when $\theta = 0$.  Hence, the method is not valid in that case. 
An implementation of this method, when $\theta > 0$, is given in Algorithm \ref{alg:GirsanovStandard} in Appendix \ref{sec:MC_Algorithms}.

We now turn back to the RNIC model.  We will denote our statistic of interest as $\E[f(\Fsimtheta(t))]$, where $\Fsimtheta(t)$ is the state of the RNIC model at time $t$ and $f: \bigcup_{m=0}^\infty \left( \ZZ^d_{\ge 0}\right)^m\to \RR$ is a function of interest.  
We will denote by $W^\theta$ the generated weight function.
Hence, we utilize the algorithms below to estimate $\frac{\partial}{\partial \theta} \E[f (\Fsimtheta(t_\text{end}))]$ by averaging the independent realizations $f(\Fsimtheta_{[\ell]}(t_{\text{end}})) W^\theta_{[\ell]}(t_{\text{end}})$, with $[\ell]$ enumerating the independent calls to the algorithm.

In fact, the details above pertaining to the Girsanov method for standard stochastic reaction networks allow us to immediately provide the Girsanov transformation method when the perturbed parameter is one of the rate constants associated with the compartment model, $\mathcal{H}_\K$.  This is because the compartment model can be simulated with no knowledge of the internal reaction networks. Specifically, the following pseudo-algorithm works:
\begin{enumerate}
\item Utilize Algorithm \ref{alg:GirsanovStandard}, the standard Girsanov method, to simulate the compartment model.  
\item In between compartment events, simulate the independently evolving reaction networks as in Step 5 of Algorithm \ref{alg:exact}.
\end{enumerate}

We turn to the case when the parameter of interest is in the reaction network $\{\I_\K\}$.  Because the reaction networks within each compartment evolve independently, we may sum the representative weight functions from the calls to $\I_{\K}$ between compartment events.  Hence, simulation in this case   is also  straightforward.  However, since this situation is different from what is standard, we provide a pseudo-algorithm.

\begin{algorithm}[Girsanov transformation method if the perturbed parameter is from the reaction network model, $\I_\K$]
\label{alg:Girsanov_RN}
Given: a stochastic reaction network $\I_{\K}$, the compartment model \eqref{eq:compartment8907}, the parameters of the model $\K$, which consist of the rate constants $\{\kappa_k\}$ for $\I_{\K}$ and the parameters of the compartment model $\{\kappa_I, \kappa_E, \kappa_F, \kappa_C\}$, initial distributions $\nu$, $\mu_0$, and $\mu$. 

All calls to random variables are independent from all others.
\begin{enumerate}[itemsep=-.05ex]
    \item Determine $M_C(0)$ via $\nu$.
    \item For each $i \in \{1, \dots, M_C(0)\}$ determine $X_i(0)$ via $\mu_0$.
    \item Set $W^\theta = 0$ and set $t=0$.
    \item Using Gillespie's Algorithm \ref{alg:Gillespie}, determine the time $\Delta$ until the next transition of the compartment model, $M_C$.
    \item If $t + \Delta \ge t_\text{end}$, set $\Delta = t_\text{end} - t$ and break after step 6.
    
    \item Using the reaction network Girsanov transformation Algorithm \ref{alg:GirsanovStandard}, for all $i \in \{1, \dots, M_C(t)\}$ simulate $X_i$ from time $t$ to time $t + \Delta$ and let $W_i$ denote the output weight function. 
    
    Set $W^\theta \leftarrow W^\theta+ \sum_{i=1}^{M_C(t)} W_i^\theta$.
    \item Determine which type of transition occurs for the compartment model at time $t + \Delta$.
    \item Update the RNIC model as detailed in the four cases listed in Section \ref{sec:RNIC} depending upon which type of transition takes place for the compartment model.
    \item Set $t \leftarrow t + \Delta$.
\end{enumerate}
Output $\Fsimtheta(t)$ and $W^\theta$.
\end{algorithm}

\subsection{Finite difference methods}
\label{sec:FD}

As detailed in Appendix  \ref{sec:MonteCarlo}, the basic idea of finite difference methods is to use the following straightforward approximation,
\begin{align}
\begin{split}
    \label{eq:768756}
\frac{d}{d\theta} \E[f(\Fsimtheta(t))] &\approx \frac{\E[f(\Fsimthetah(t)) - f(\Fsimtheta(t))]}{h}\\
&\approx \frac{1}{n} \sum_{\ell=1}^n\frac{1}{h} \left( f(\Fsimthetah_{[\ell]}(t)) - f(\Fsimtheta_{[\ell]}(t))\right),
\end{split}
\end{align}
(or a centered difference) where the key fact is that $\Fsimthetah_{[\ell]}$ and $\Fsimtheta_{[\ell]}$ are generated on the same probability space.  Specifically, we want $\Fsimthetah_{[\ell]}$ and $\Fsimtheta_{[\ell]}$ to be coupled so that 
\[
\text{Var}\left( f(\Fsimthetah_{[\ell]}(t)) - f(\Fsimtheta_{[\ell]}(t))\right)
\]
 is reduced.\footnote{In the appendix we introduced the centered finite difference, whereas here we are using the forward difference.  The variance is not affected by this choice and so we utilize the less notationally cumbersome choice here.}

We will use five basic coupling methods, each with different ``flavors'' depending on what type of parameter is being perturbed. 
We begin with a simple introduction to each.

\vspace{.1in}

\noindent \textbf{Method 1: Independent Samples}.  This is straightforward.  With this method, the processes are generated independently.  This  should be viewed as the base case. 
\hfill $\ddagger$

\vspace{.1in}

\noindent \textbf{Method 2: Common Random Variables (CRV)}.  This is nearly as simply as generating the processes independently and consists of simply reusing the seed of the random number generator for the calls to the different processes.  Alternatively, and equivalently, 
vectors of uniform$[0,1]$---or other---random variables that can be used to simulate the process can be pre-generated. This is simply an implementation choice and not a theoretical constraint. \hfill $\ddagger$

\vspace{.1in}

\noindent \textbf{Method 3: Common Reaction Path  (CRP)}.  This method was developed for standard stochastic reaction networks, so we  explain it in that language.  Hence, consider a stochastic reaction network with jump directions $\{\zeta_k\}$ and intensity functions $\{\lambda_k^\theta\}$, which are parameterized by $\theta$.  The Common Reaction Path method for such a process couples $X^\theta$ and $X^{\theta + h}$  by sharing the unit-rate Poisson processes in \eqref{eq:parameterized_RTC}, see \cite{Khammash2010}.  That is, $X^\theta$ and $X^{\theta+h}$ are coupled (generated) in the following manner,
\begin{align*}
X^{\theta+h}(t) &= X^{\theta+h}(0) + \sum_k Y_k \left( \int_0^t \lambda_k^{\theta+h} (X^{\theta+h}(s)) ds \right) \zeta_k\\
X^\theta(t) &= X^\theta(0) + \sum_k Y_k \left( \int_0^t \lambda_k^\theta (X^\theta(s)) ds \right) \zeta_k,
\end{align*}
where the key point is that the same unit-rate Poisson processes, $\{Y_k\}$, are utilized for both $X^\theta$ and $X^{\theta+h}$.  We will often refer to this as the classical CRP method in order to differentiate it from the local-CRP method introduced next.  \hfill $\ddagger$

\vspace{.1in}

\noindent \textbf{Method 4: local Common Reaction Path  (local-CRP)}.  We again explain this method in the context of standard stochastic reaction networks, using the same notation as above (i.e., jump directions $\{\zeta_k\}$ and intensity functions $\{\lambda_k^\theta\}$).  
The idea of the local-CRP coupling, as introduced in \cite{anderson2015asymptotic}, is to discretize your time interval $[0,T]$ into multiple sub-intervals and for each such subinterval  generate the coupled processes using a new set of independent unit-rate Poisson processes.   That is, use the classical CRP coupling mentioned above within each sub-interval, with new Poisson processes.

More formally, we begin by  discretizing an interval of time $[0,T]$ by letting $\pi = \{0 = s_0<s_1<\cdots < s_r = T\}$.  Then, let $\{Y_{km},  \text{ for } k \in \{1, \dots, R\}, \ m \in\{ 0, 1, 2,\dots\}\}$ be a set of independent, unit-rate Poisson processes, where $R$ is the number of reactions (i.e., the size of the set $\mathcal{R}$).  Then, the local-CRP coupling over $[0,T]$ with respect to $\pi$ is the solution to

\begin{align*}
X^{\theta+h}(t) &= X^{\theta+h}(0) + \sum_{k=1}^R \sum_{m=0}^\infty Y_{km} \left( \int_{t\wedge s_m}^{t\wedge s_{m+1}} \lambda_k^{\theta+h} (X^{\theta+h}(s)) ds \right) \zeta_k\\
X^\theta(t) &= X^\theta(0) + \sum_{k=1}^R\sum_{m=0}^\infty Y_{km} \left(  \int_{t\wedge s_m}^{t\wedge s_{m+1}}  \lambda_k^\theta (X^\theta(s)) ds \right) \zeta_k.
\end{align*}
We note here that in the implementations utilized in this paper, the discretization time points $\{s_i\}$ will themselves be random.\hfill $\ddagger$

\vspace{.1in}

\noindent \textbf{Method 5:  Split Coupling (SC)}.  As with the two methods above,  we again explain this method in the context of standard stochastic reaction networks, using the same notation as above (i.e., jump directions $\{\zeta_k\}$ and intensity functions $\{\lambda_k^\theta\}$).   The Split Coupling method for such a process couples $X^\theta$ and $X^{\theta + h}$ by constructing them in the following manner \cite{AndCFD2012,AndHigham2012}, 
\begin{align}
\label{eq:5467890}
\begin{split}
X^{\theta+h}(t) &= X^{\theta+h}(0) + \sum_k Y_{k,1} \left( \int_0^t \min\{ \lambda_k^{\theta+h} (X^{\theta+h}(s)), \lambda_k^\theta (X^\theta(s))\} ds \right) \zeta_k\\
&\qquad+ \sum_k Y_{k,2} \left( \int_0^t\lambda_k^{\theta+h} (X^{\theta+h}(s))- \min\{ \lambda_k^{\theta+h} (X^{\theta+h}(s)), \lambda_k^\theta (X^\theta(s))\} ds \right) \zeta_k\\
X^\theta(t) &= X^\theta(0) + \sum_k Y_{k,1} \left( \int_0^t \min\{ \lambda_k^{\theta+h} (X^{\theta+h}(s)), \lambda_k^\theta (X^\theta(s))\} ds \right) \zeta_k\\
&\qquad+\sum_k Y_{k,3} \left( \int_0^t\lambda_k^{\theta} (X^{\theta}(s))- \min\{ \lambda_k^{\theta+h} (X^{\theta+h}(s)), \lambda_k^\theta (X^\theta(s))\} ds \right) \zeta_k,
\end{split}
\end{align}
where $\{Y_{k,1},Y_{k,2},Y_{k,3}\}$ are all independent unit-rate Poisson processes.  Note that both processes share the jump processes associated with the Poisson processes $\{Y_{k,1}\}$. For the purposes of this paper, we will simply note that the Split Coupling works (in that $X^\theta$ and $X^{\theta+h}$ have the correct distributions) because of the fact that Poisson processes are additive in that if $Y_1, Y_2,$ and $Y_3$ are unit-rate Poisson processes, $Y_1$ and $Y_2$ are independent, and if $\lambda_1,\lambda_2$  are integrable functions, then $Y_1(\int_0^t \lambda_1(s)ds) + Y_2(\int_0^t \lambda_2(s))$ is equal in distribution to $Y_3(\int_0^t \left( \lambda_1(s) +\lambda_2(s) \right) ds)$.
\hfill $\ddagger$

\vspace{.1in}

The question now is how we implement these basic methods in the RNIC setting. There are two distinct cases to consider: whether the parameter being perturbed is a compartment parameter ($\kappa_I,\kappa_E,\kappa_F$, or $\kappa_C$) or is a parameter from the reaction network (e.g., a rate constant for stochastic mass-action kinetics).  One case is demonstrably more straightforward, and so we start with that.

\vspace{.1in}

\noindent \textbf{Case 1: $\theta$ is a parameter of the reaction network, $\I$}.

\vspace{.05in}

In this case
the compartment model \eqref{eq:compartment8907}, whose transitions and rates do not depend upon the states of the chemical systems, can be shared between $\Fsimtheta$ and $\Fsimthetah$.  Hence, only a slight modification to Algorithm \ref{alg:exact} is needed.  We provide that modification here assuming a Gillespie implementation for the simulation of the compartments.  We also utilize  only the local-CRP below as opposed to the classical CRP.

\begin{algorithm}[Generic coupling for an RNIC model when perturbed parameter is from the reaction network, $\I$]
\label{alg:coupling_RN}
Given: a reaction network $\I$ with jump directions $\{\zeta_k\}$ and intensity functions $\lambda_k^\theta$, the compartment model \eqref{eq:compartment8907} with parameters  $\{\kappa_I, \kappa_E, \kappa_F, \kappa_C\}$, initial distributions $\nu$, $\mu_0$, and $\mu$. 

All calls to random variables are independent from all others.
\begin{enumerate}[itemsep=-.05ex]
    \item Determine $M_C(0)$ via $\nu$.
    \item For each $i \in \{1, \dots, M_C(0)\}$ determine $X_i^\theta(0)$ via $\mu_0$ and set $X_i^{\theta+h}(0) = X_i^\theta(0).$
    \item Set $t = 0$.
    \item Using Gillespie's Algorithm \ref{alg:Gillespie}, determine the time $\Delta$ until the next transition of the compartment model, $M_C$.
    \item If $t + \Delta \ge t_\text{end}$, set $\Delta = t_\text{end} - t$ and break after step 6.
    
    \item Using one of independent samples, the CRV coupling, the local-CRP coupling or the Split Coupling, for each $i \in \{1, \dots, M_C(t)\}$ simulate $X_i^\theta$ and $X_i^{\theta + h}$ from time $t$ to time $t + \Delta$.   Note the following.
    \begin{itemize}
    \item For independent realizations, this step amounts to making two independent calls to an exact simulation method, as can be found in Appendix \ref{sec:MC_Algorithms}.

    \item For the CRV method, this step consists of generating a new seed for each compartment and using that seed for both calls to the exact simulation method. Equivalently, one can simply pass a given vector of random variables to each of the two calls of an exact simulation method, and use those random variables to construct the processes.  We chose the latter option of passing a vector of random variables and note that this is an implementation choice, as opposed to a theoretical constraint.
    
    \item For the local-CRP method, this step consists of generating new unit-rate Poisson processes for each reaction type in each compartment (via a sequence of independent unit-rate exponentials) and utilizing those Poisson processes to separately construct both $X_i^\theta$ and $X_i^{\theta+h}$ on $[t,t+\Delta]$.

    \item For the Split Coupling, this step simply consists of simulating \eqref{eq:5467890} within each compartment.  This can be done in a straightforward manner via Gillespie's algorithm (since \eqref{eq:5467890} is also a Markov process) or the next reaction method \cite{AndCFD2012,AndHigham2012}.
    \end{itemize}
    
    \item Determine which type of transition occurs for the compartment model at time $t + \Delta$.
    \item Update the RNIC model as detailed in the four cases listed in Section \ref{sec:RNIC} depending upon which type of transition takes place for the compartment model.  (See below for more details.)
    \item Set $t \leftarrow t + \Delta$, and return to step 4.
\end{enumerate}
Output $\Fsimtheta(t_{\text{end}})$ and $\Fsimthetah(t_\text{end})$.
\end{algorithm}

\begin{remark}
    \label{remark:fragmentation}
It is worth delving a bit into step 8 above in order to ensure that the coupling is as tight as possible, with special focus on when a fragmentation event takes place.  We do the following for all of our simulations.  There are, of course, four cases: one for each of the possible compartment events.
\begin{enumerate}
    \item If the transition of the compartment model is an arrival, then the new compartment is initialized via $\mu$ once and that value is assigned to both the $\theta$ process and the $\theta+h$ process. 
    
    \item If the transition of the compartment model is departure/exit, then the compartment deleted is the same for both processes.
    
    \item If the transition of the compartment model is a coagulation (merger), then the two compartments that merge are the same for both the $\theta$ and $\theta+h$ process.  Moreover, the choice of which of the two is deleted (and which collects the material) is also the same.
    
    \item If the transition of the compartment model is a fragmentation, then the compartment chosen for splitting is the same for both processes.  However, how the contents of the chosen compartment are divided  should be considered carefully.  For example, in our single-path simulations our splitting rule is the following: each molecule of each species will choose its compartment via a fair coin flip.  Therefore, when generating a single path (i.e., no coupling) we may simply do the following: for each species, generate a binomial random variable with parameters $m=\#$ of that species present in the chosen compartment and $p= 1/2$.  However, for our coupled processes it is possible (and even likely) that the compartment chosen will have different numbers of each species for the $\theta$ and $\theta+h$ processes.  Because of this, we suggest (and carry out) the following.  
    
    Suppose it is the $i$th compartment chosen for fragmentation and the $j$th species has abundances $X^\theta_{i,j}$ and $X^{\theta + h}_{i,j}$, respectively.  Suppose further that $X^\theta_{i,j} \le X^{\theta + h}_{i,j}$ (with a symmetric construction in the other case).  Then, for the $\theta$ process, we choose a binomial random variable with parameters $m = X^\theta_{i,j}$ and $p = 1/2$.  Call that value $M_1$.  For the $\theta+h$ process, we choose a binomial random variable with parameters $M_2 = X^{\theta + h}_{i,j}-X^\theta_{i,j}$ and $p = 1/2$, and then add that value to $M_1$.  We carry out that basic idea for each $j \in \{1,\dots, d\}$.  In this manner, we have coupled the splitting (and therefore hopefully the processes) as tightly as possible.\hfill $\square$
\end{enumerate}
\end{remark}

We make one  further comment  regarding the use of the local-CRP method versus the classical CRP method.  We noted explicitly above that when the local-CRP method is being used, new unit-rate Poisson processes are to be utilized for each reaction network between each transition of the compartment model.  This is done for practical purposes: since the compartments are constantly merging, fragmenting, etc., if the classical CRP method were to be used it would be unclear which Poisson processes go with which compartment.  Hence, the local-CRP is more natural to use in this setting.  In  \cite{anderson2015asymptotic}, it was proven that as the number of replacements of the Poisson processes increases, the coupled processes converge in distribution to the coupled processes generated via the Split Coupling.  We will see this play out in our examples of Section \ref{sec:examples} and it explains why the variance of the local-CRP method, as we have implemented it, closely matches that of the Split Coupling in the RNIC setting.

\vspace{.1in}

\noindent \textbf{Case 2: $\theta$ is a parameter of the compartment model $\mathcal{H}$}.

\vspace{.05in}

Now the compartment processes $M_C^\theta$ and $M_C^{\theta+h}$ eventually diverge  and  the two processes $\Fsimtheta$ and $\Fsimthetah$ can no longer share the compartment model.  We cover the various coupling strategies in this case.

\vspace{.1in}

\noindent \textbf{Independent samples.}  We generate the process $\Fsimtheta$ via Algorithm \ref{alg:exact}, change the parameter, and generate the process $\Fsimthetah$ via Algorithm \ref{alg:exact}. The processes are constructed independently.

\vspace{.1in}

\noindent \textbf{Common Random Variables.}  We fix a seed of the random number generator we have not used as of yet. 
We generate the process $\Fsimtheta$ via Algorithm \ref{alg:exact}, change the parameter, and generate the process $\Fsimthetah$ via Algorithm \ref{alg:exact} using the same seed.

\vspace{.1in}

\noindent \textbf{Classical CRP}. We only couple the processes $\Fsimtheta$ and $\Fsimthetah$ via the compartments (utilizing the classical CRP method of \cite{Khammash2010}) and the compartment events.  We do not couple the stochastic reaction networks taking place within the compartments between the events (see Remark \ref{rmk:blahblah} for why).  Note that to couple the processes as tightly as possible, we need to send multiple vectors of random variables for the generation of our two processes: one or more vectors for each possible compartment event that can take place, as well as the needed Poisson processes for the compartment model.  Specifically, we generate the following random variables and processes and use them to generate both $\Fsimtheta$ and $\Fsimthetah$:

\begin{itemize}[itemsep=-.05ex]
    \item The Poisson processes $Y_I,Y_E,Y_C,Y_F$ (generated as a vector of unit-exponentials);
    \item A vector of uniform[0,1] random variables to determine the initial values of the compartments (both at time 0 and those that arrive);
    \item A vector of uniform[0,1] random variables to determine which compartment is chosen for deletions;
    \item A vector of uniform[0,1] random variables to determine which compartments coagulate;
    \item A vector of uniform[0,1] random variables to determine which compartment fragments;
     \item A vector of uniform[0,1] random variables to determine the needed binomial random variables for a fragmentation event.
\end{itemize}
These vectors of random variables are utilized in the obvious way, but we point to the freely available Matlab code for precise details.  

\vspace{.1in}

\noindent \textbf{Split Coupling.}   We couple the compartment models $M_C^\theta$ and $M_C^{\theta+h}$ using the split coupling.  
Events that take place using the split coupling can be ``shared'' (which occurs when one of the counting processes associated to $Y_{k,1}$ in \eqref{eq:5467890} takes place--these have intensity functions that are the minimum of the two individual intensity functions) or not.  If the event is not shared, then one simply updates the relevant compartment model (either $M_C^\theta$ or $M_C^{\theta+h}$) as usual.  If the event is shared, then the following procedures are followed:
\begin{enumerate}
    \item If the event is an arrival, both new compartments are initialized from $\mu$ with the same value.
    
    \item If the event is a deletion, the same uniform[0,1] random variable  is used to determine which compartment is deleted.  Note that this does not necessarily mean the processes will delete the same numbered compartment.

    \item If the event is a coagulation, the same uniform[0,1] random variables are used to determine which compartments merge for the two processes.

    \item If the event is a fragmentation, the same uniform[0,1] random variable  is used to determine which compartment fragments.  Moreover, we utilize the same procedure for fragmentation that was detailed in point 4 of Remark \ref{remark:fragmentation}.

\end{enumerate}
We point the reader to the freely available Matlab code for the specific implementation.

When using the split coupling for the compartment model, compartment events take place simultaneously for the two processes $\Fsimtheta$ and $\Fsimthetah$.  Hence, we are also able to use the split coupling for the chemical processes, $\{X_i^\theta\}$ and $\{X_i^{\theta+h}\}$ between such events (as in the case when a chemical parameter is being perturbed).  Specifically, we  couple the chemical processes sequentially as much as possible.  That is, we generate 
\[
(X_1^\theta,X_1^{\theta+h}), \dots (X_m^\theta,X_m^{\theta+h}),\quad \text{ where } \quad m = \min\{M_C^\theta, M_C^{\theta+h}\},
\]
using the split coupling.  Then, for any other compartments (for example, if $M_C^\theta > M_C^{\theta+h}$) we simply generate those paths independently.

We will also later provide results for when we use the Split Coupling on the compartment model (as detailed above) but \textit{not} for the chemical models.  In that case, we generate the processes $\{X_i^\theta\}$ and $\{X_i^{\theta+h}\}$ independently.  We do this solely to compare with the CRP method (in which coupling of the chemical models is also not done).

\begin{remark}
\label{rmk:blahblah}

Note that the local CRP method was not included in the list of coupling strategies used in this case. The reasons for this are as follows.
Because $\theta$ is assumed to be a compartment parameter, the timing of the compartment events of the processes $\Fsimtheta$ and $\Fsimthetah$ will not  necessarily be equal.  In fact, after the first compartment event that is associated with the perturbed parameter takes place for one of $\Fsimtheta$ or $\Fsimthetah$, the  two compartment processes do not have any simultaneous compartment events for exits, fragmentations, or coagulations ever again (they will always share input events, since that rate is state-independent).   Moreover, because the processes $\Fsimtheta$ and $\Fsimthetah$ are generated separately there is no way for the process $\Fsimtheta$ to ``know'' when there is a compartment event for $\Fsimthetah$, and vice-versa.  Combining the above two points, there is no way for either process to ``know'' when a compartment event for the other took place, and so there is no way to update the Poisson processes for the chemical system at those times in a shared manner.  
Using the language of stochastic processes, the issue being pointed out here stems from the fact that while the CRP method produces two paths, those coupled paths are not Markovian.

Theoretically, a way around this problem would be to generate the compartment models for $\Fsimtheta$ and $\Fsimthetah$ simultaneously from $Y_I, Y_E, Y_F$, and $Y_C$, while tracking all information used by each process.  However, this would entail a large infrastructure in order to keep track of how much of the various Poisson processes $Y_I, Y_E, Y_F$, and $Y_C$ each of $\Fsimtheta$ and $\Fsimthetah$ has seen by a given time and various vectors accounting for  ``future'' information.  This would be quite cumbersome and, potentially, quite slow.  Moreover, we know from the analysis \cite{anderson2015asymptotic} that the resulting variance would be nearly identical to that of the Split Coupling (which does not require this added machinery).  Hence, we choose not to perform this implementation. \hfill $\square$
\end{remark}

\section{Examples}
\label{sec:examples}

We provide two RNIC models as test cases for the various methods outlined above.  For each model, we  perturb both a chemical parameter and a compartment parameter. For the second model we also consider the situation in which more than one parameter is perturbed.  
All of the computations reported in this section were performed using Matlab version 2024b on a 2019 MacBook Pro using a 2.3 GHz 8-Core Intel I9 processor. 

\begin{example}
\label{example:BD}
\textbf{Birth and death}.  Our first model is  one of the simplest  RNIC models, though it presents a nice test case that is also analytically tractable.  The model consists of a chemical system that is a birth and death process (termed an $M/M/\infty$ queue in the queueing literature) and a compartment model that is also a birth and death process.
Specifically, we have the following:
\begin{equation*}
\text{(Chemistry)} \quad \emptyset \overset{\kappa_b}{\underset{\kappa_d}{\rightleftharpoons}} S,\qquad  \text{(Compartment)} \quad \emptyset \overset{\kappa_I}{\underset{\kappa_E}{\rightleftharpoons}}  C, \qquad \text{(Initial distribution)}\quad  \text{Poisson$(\phi)$}.
\end{equation*}
Hence, both the fragmentation $(\kappa_F)$ and coagulation $(\kappa_C)$ parameters for the compartment model are  set to zero, and so the compartments themselves do not interact.  This model was considered as an example RNIC model in  \cite{DZ} and the chemical portion was utilized as an example in \cite{AndCFD2012}, where the split coupling was introduced in the context of parametric sensitivity analysis.
Following \cite{DZ}, we will take the rate constants for the chemical model to be $\kappa_b = 1$ and $\kappa_d = 0.1$ and for the compartment model to be $\kappa_I = 1$ and $\kappa_E = 0.01.$ Also following \cite{DZ}, we set $\phi = 10$, so that compartments arrive in equilibrium \cite{AndProdForm}.  We initialize the RNIC model itself with zero compartments.    Denote by $S_{tot}(t)$ the total number of $S$ molecules summed over the compartments at time $t$. We will estimate
\[
\frac{d}{d\kappa_b} \E[S_{tot}(t)] \quad \text{ and } \quad \frac{d}{d\kappa_E} \E[S_{tot}(t)].
\]

This model lends itself to analysis, which makes it particularly useful in our study.  By \cite{anderson2016product}, we know that the stationary distribution is Poisson with mean 100.  In fact, more can be determined.  Denote by $\bar c(t)$ the mean value for the number of compartments at time $t \ge 0$.  By \cite{anderson2020time}, the  distribution of the compartment process at time $t\ge 0$ is Poisson with a mean given by the solution to the ODE $\frac{d}{dt} \bar c(t) = \kappa_I - \kappa_E \bar c(t)$.  
 By equation (19) in \cite{DZ} the equations governing the means of the compartments and the total amount of $S$ are
\begin{align}
\begin{split}
\label{eq:means}
	\frac{d}{dt}  \bar c(t) &= \kappa_I - \kappa_E \bar c(t)\\
	\frac{d}{dt}  \E[S_{tot}(t)] &= \kappa_I \phi - \kappa_E \E[S_{tot}(t)] + \kappa_b \bar c(t) - \kappa_d  \E[S_{tot}(t)].
	\end{split}
\end{align}
These linear ODEs can be solved quite easily and so can be used to provide us with the exact values of the sensitivities and, importantly for us, the precise biases of the finite difference methods.  We do not provide the solutions here explicitly as they are trivial to compute with any software package (e.g., Maple, Mathematica, etc.,) but take up quite a bit of space.

\vspace{6pt}

\noindent \textbf{Estimation of $\frac{d}{d \kappa_b} \E[S_{tot}(t)]$}.  Since $\kappa_b$ is a parameter of the chemical model, we utilize Algorithm \ref{alg:Girsanov_RN} for the Girsanov method and Algorithm \ref{alg:coupling_RN} for the finite difference methods. 
The solution to $\E[S_{tot}(t)]$ given by the equations \eqref{eq:means} has the general form $g_1(t) + \kappa_b g_2(t)$, for functions $g_1$ and $g_2$, which depend on rate constants other than $\kappa_b$.  Hence, the dependence is linear in $\kappa_b$ and the finite difference methods are \textit{unbiased} estimators for $\frac{d}{d \kappa_b}  \E[S_{tot}(t)]$. Thus, only the variances of the different methods need to be considered.  Moreover, because the finite difference methods are unbiased, we chose to utilize the forward difference as opposed to the centered difference (see Appendix \ref{sec:MonteCarlo}), but this choice is immaterial.

For each of the finite difference methods, we utilized $n=$ 1,000 paths to estimate the derivative.  For the finite difference methods, each ``path'' consists of a realization of the coupled processes $(\Fsimthetah,\Fsimtheta)$ (therefore, we end with 1,000 copies of each of $\Fsimtheta$ and $\Fsimthetah$). Moreover, we used a perturbation of $h = 1/10$, which is 10\% of the parameter value.    For the Girsanov method, we utilized 2,000 paths, which yields a similar computational budget to the finite difference methods. 

\begin{figure}
\begin{center}
\subfigure[Girsanov transformation method]{\includegraphics[width=0.49\textwidth]{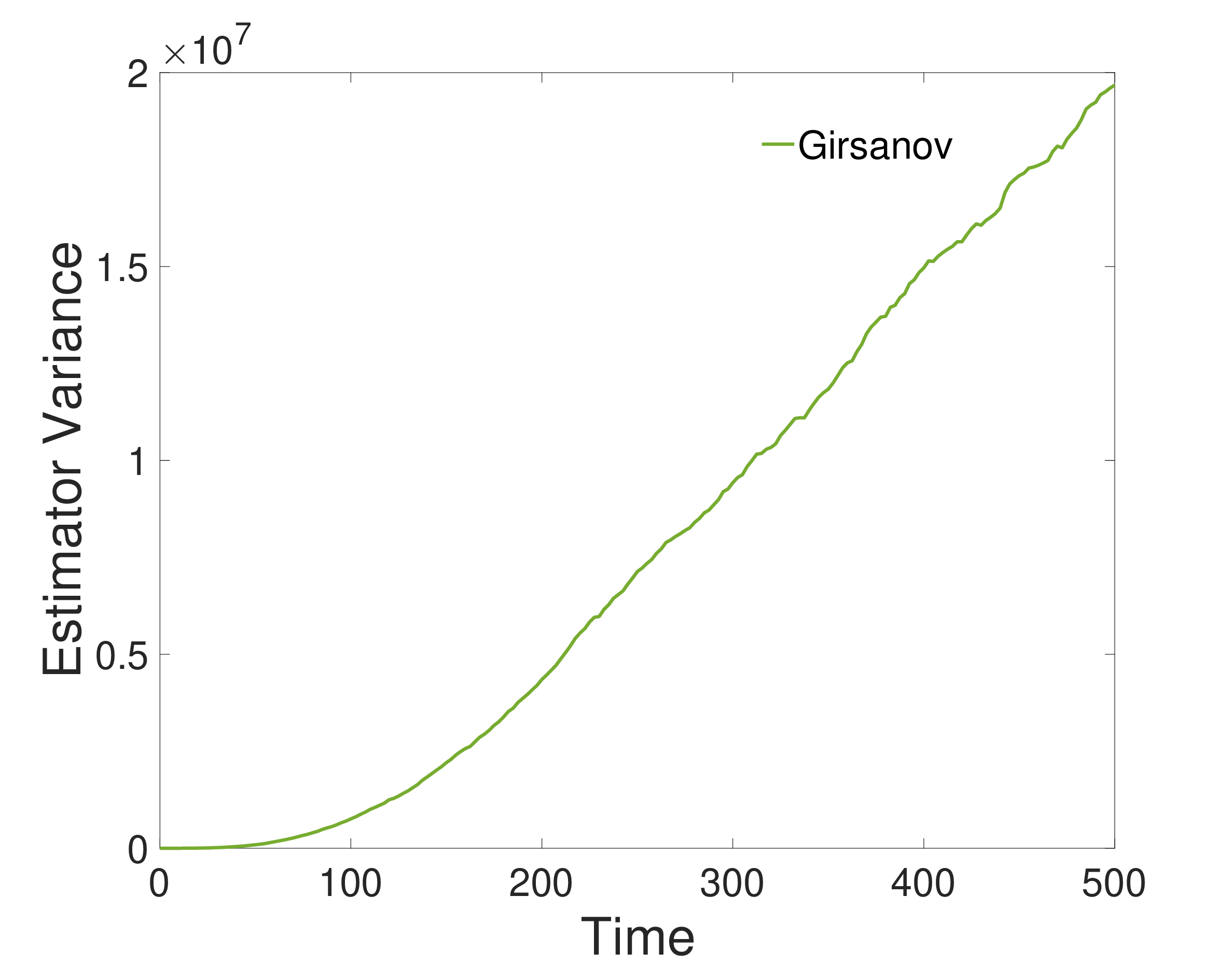}}
\subfigure[Compartment model not shared]{\includegraphics[width=0.49\textwidth]{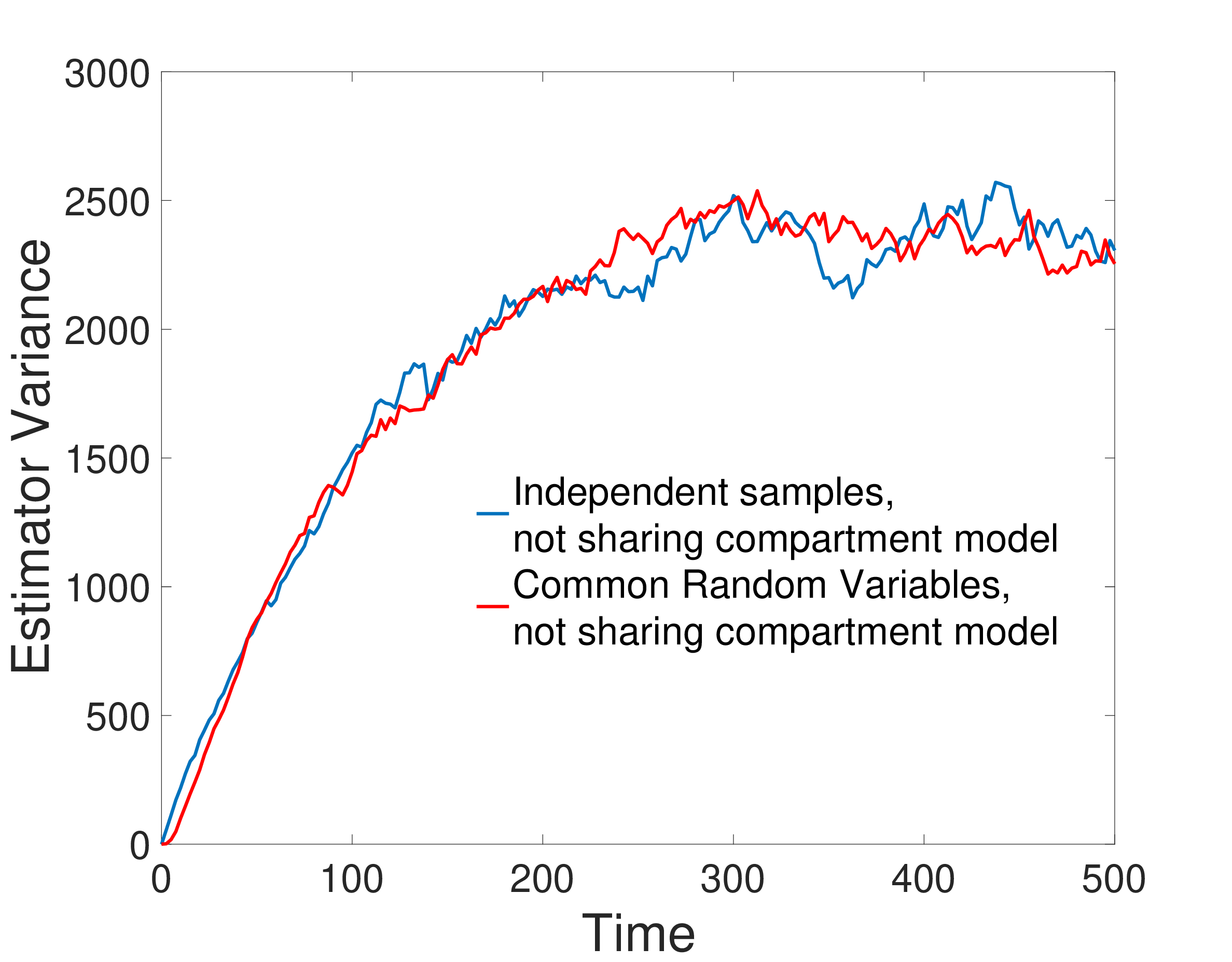}}
\subfigure[Compartment model shared]{\includegraphics[width=0.49 \textwidth]{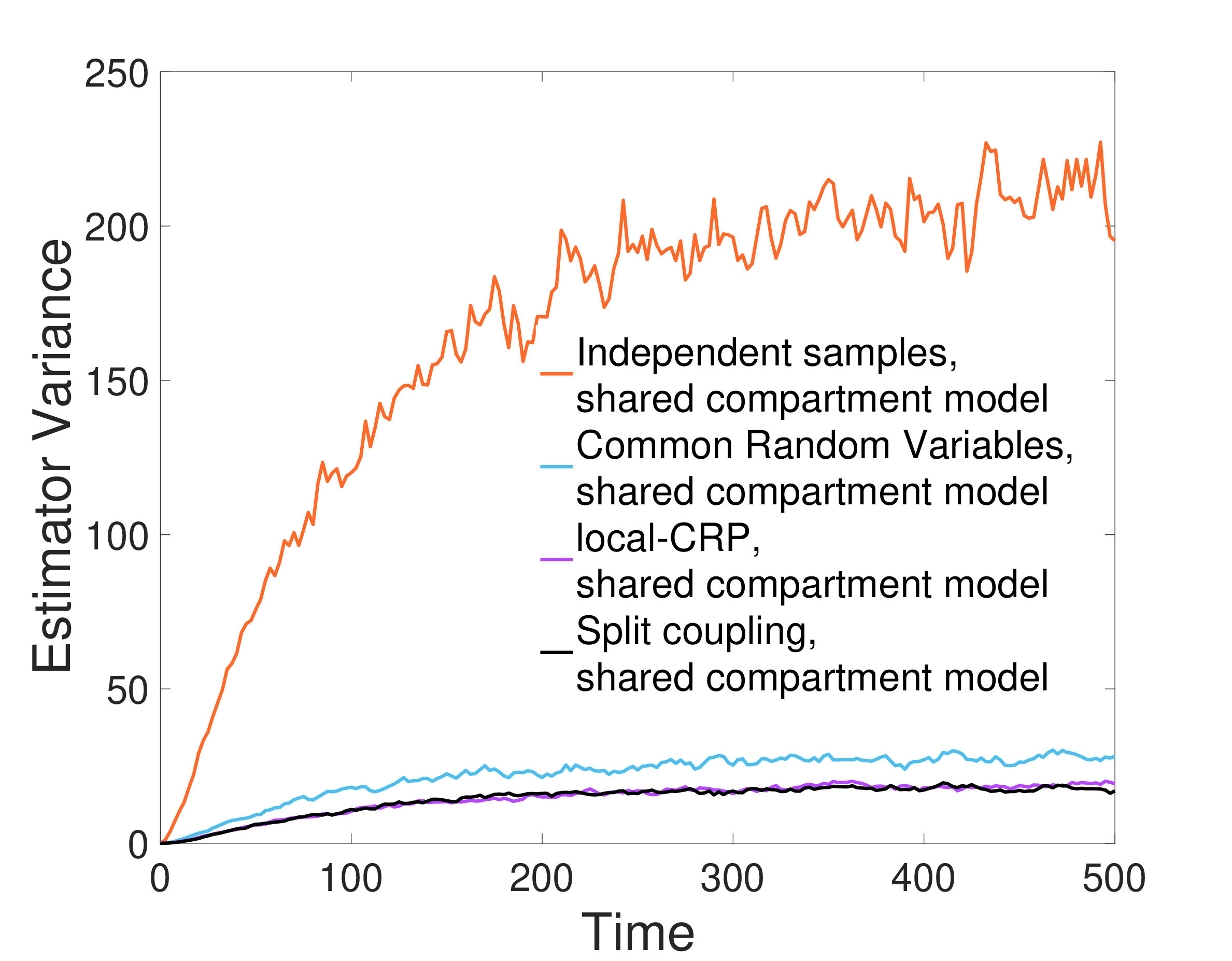}}
\end{center}
\caption{Estimator variances \eqref{eq:estimatorvarianceformula} for  $\frac{d}{d\kappa_b} \E[ S_{tot}(t)]$ using various methods.  For the (forward) finite difference methods, we used a perturbation of $h = 1/10$ and $n=$ 1,000 coupled paths.  For the Girsanov method we used $n =$ 2000 paths.  Note the vastly different scales on the $y$-axis.}
\label{fig:786yuoihjl}
\end{figure}

In  Figure \ref{fig:786yuoihjl} we report the estimator variances
\begin{align}
    \label{eq:estimatorvarianceformula}
    \frac{\text{Var}\left(f(\Fsimtheta(t))\cdot W(t)\right)}{n} \quad \text{ and } \quad \frac{\text{Var}\left(f(\Fsimthetah(t))-f(\Fsimtheta(t))\right)}{hn},
\end{align}
for the Girsanov and finite difference methods, respectively,
where $f$ is the function that returns ``Total S'', or $S_{tot}$, and $W$ is the weight function for the Girsanov method. We explicitly point out that the Split Coupling and the local-CRP method have nearly identical estimator variances (as predicted from \cite{anderson2015asymptotic}), and both variances are substantially lower than the other methods.
We also note that, for the sake of comparison, we have provided two methods not detailed in the previous section: the use of independent samples and Common Random Variables that do not share the compartment model (making these the most naive methods possible).  We do so to be able to visualize the benefit of coupling through the compartments, which is substantial.

We provide in Table \ref{table:9087} the time required to generate the paths for each of the methods.  Since the time required for each method depends intimately on the hardware and software used and on the particulars of the implementation, it is sometimes difficult to interpret and could possibly be improved, perhaps dramatically, by clever implementations.  Hence, we also provide in Table \ref{table:9087} the number of random variables used in the whole computation (but we do \textit{not} include those random variables generated but not used by the computation in the local-CRP method).  The number of random variables used is largely independent of implementation and can be taken as a  (rough) proxy for the overall computational complexity. 
Figure \ref{fig:786yuoihjl} and Table \ref{table:9087} together paint a clear picture: for this particular problem, the Split Coupling is the most efficient.

\begin{table}
    \centering
    \begin{tabular}{|l|c|c|}\hline
        \hspace{2in}Method & Time & RVs used \\
        \hline
        Girsanov & 128 seconds & $5.1\times 10^8$ \\\hline
        Independent samples, not sharing compartment model & 129 seconds & $5.2 \times 10^8$ \\\hline
        Common Random Variables, not sharing compartment model & 126 seconds & $5.2 \times 10^8$  \\\hline
        Independent samples, shared compartment model & 111 seconds & $5.2 \times 10^8$ \\\hline
        Common Random Variables, shared compartment model & 167 seconds & $5.2\times 10^8$  \\\hline
        local-CRP, shared compartment model & 299 seconds & $5.3\times 10^8$ \\\hline
        Split Coupling, shared compartment model & 118 seconds & $2.7\times 10^8$ \\\hline
    \end{tabular}
    \caption{Simulation times for the various methods for the estimation of $\frac{d}{d\kappa_b}\E[S_{tot}(t)]$.    $n=$ 1,000 coupled paths were used for the finite difference methods and $n=$ 2,000 paths were used for the Girsanov method.}
\label{table:9087}
\end{table}

\vspace{6pt}

\noindent \textbf{Estimation of $\frac{d}{d\kappa_E}  \E[S_{tot}(t)]$.} 
 As $\kappa_E$ is a parameter of the compartment model, there is no longer the possibility of sharing the compartment model.  Hence, we  utilize the classical CRP method instead of the local-CRP method.

 We may again utilize the equations \eqref{eq:means} to calculate the desired sensitivity and the biases for the finite difference methods.  Over the time frame of interest, $t \in [0,500]$, the sensitivity starts at zero and decreases to -90,000 (image not shown here, but is easily generated with Maple, Mathematica, or Matlab).  If we use a perturbation of $h = 1/1000$, which is again 10\% of the nominal parameter value $\kappa_E = 1/100$, with a forward finite difference, then the bias ranges from zero (at time $t = 0$) to around 8,000 (at time $t = 500$).  A centered difference provides a bias that ranges from 0 to approximately 180.  This is a good approximation and so we choose that value for $h$ combined with the centered finite difference.

 We once again simulated $n=$ 1,000 paths of the processes in the case of finite differences and $n=$ 2,000 paths for the Girsanov method.  See Figure \ref{fig:birthanddeath_comp} for the estimator variances of the various methods.  The Split Coupling provides, by far, the lowest variance.  The CRP method required 155 CPU seconds and $4.9\times 10^8$ random variables to make the  computation.  The times required and numbers of random variables used for  the other methods were  all essentially the same to those in the previous case (for the estimation of $\frac{d}{d\kappa_b}  \E[S_{tot}(t)])$) and reported in Table \ref{table:9087}.  Hence, we do not bother to reproduce those numbers. The resulting estimator variance for the Split Coupling maxes out at around $3.5\times 10^5$, yielding an estimator standard deviation of around 591, which is a reasonable value given a bias of 180.  \hfill $\square$


\begin{figure}
\begin{center}
\subfigure[Girsanov transformation method]{\includegraphics[width=0.49\textwidth]{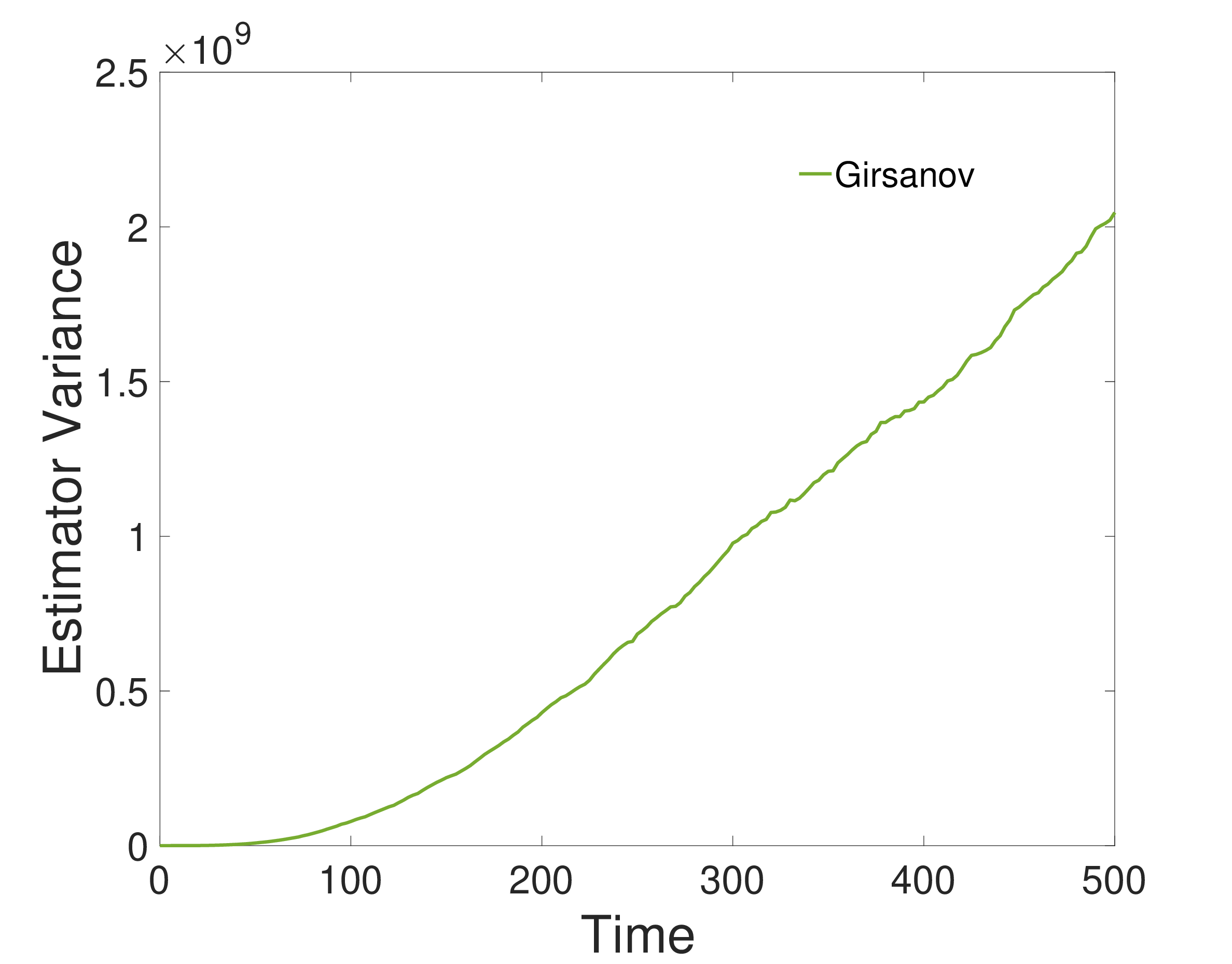}}
\subfigure[Other methods]{\includegraphics[width=0.49\textwidth]{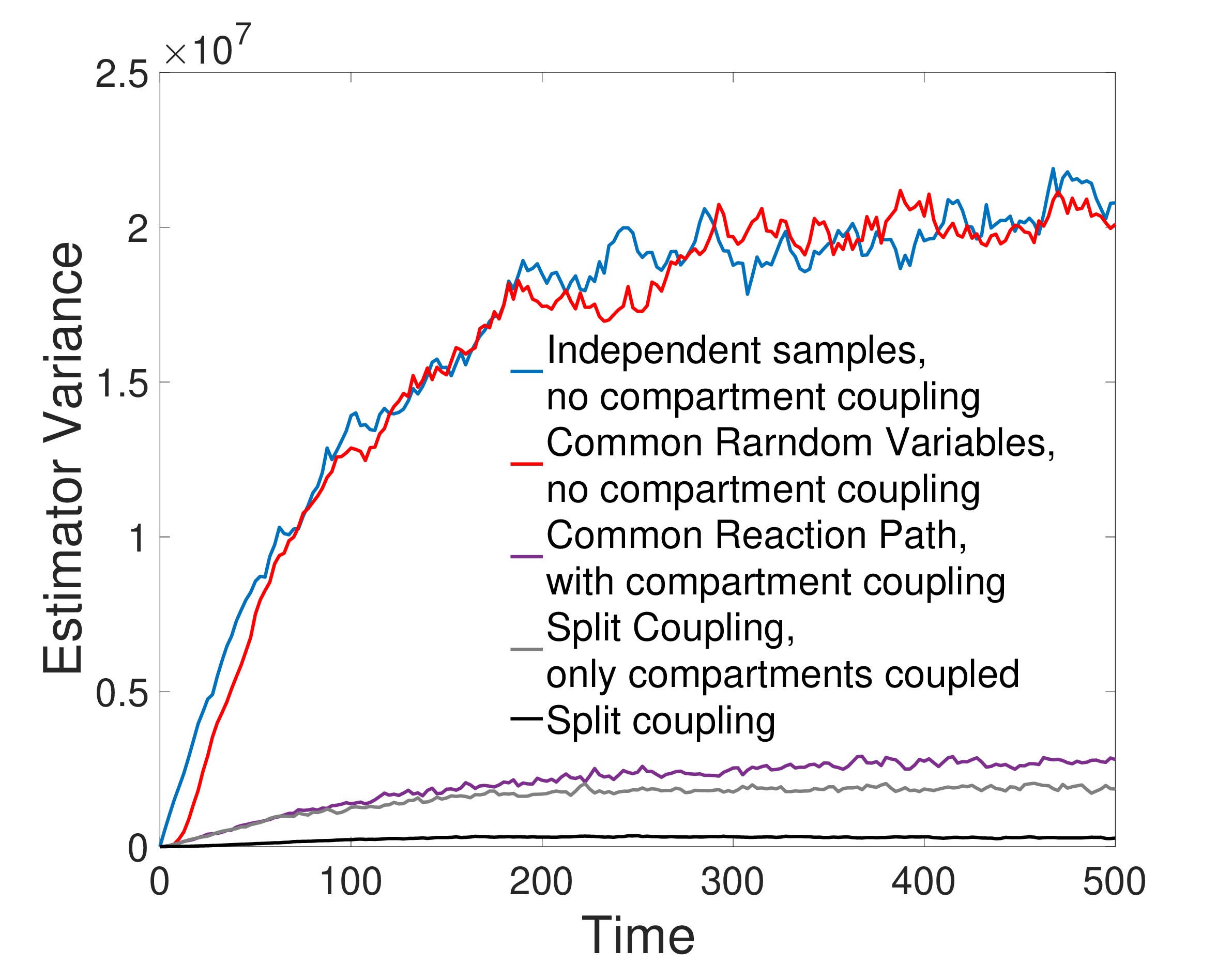}}
\end{center}
\caption{Estimator variances \eqref{eq:estimatorvarianceformula} for  $\frac{d}{d\kappa_E} \E[S_{tot}(t)]$ using various methods. For the (centered) finite difference methods, we used a perturbation of $h = 1/1000$ and $n=$ 1,000 coupled paths.  For the Girsanov method we used $n =$ 2000 paths.  Note that the Split Coupling, with both the compartment model and chemistry model being coupled, as detailed at the end of Section \ref{sec:FD}, has by far the lowest variance and is also the fastest (and utilizes the fewest random variables).}
\label{fig:birthanddeath_comp}
\end{figure}
\end{example}

\begin{example}[Genetic model with coagulation]
\label{example:gene}
Our second example has a chemistry that consists of transcription, translation, and dimerization of the resulting protein.  This particular model  is  Example 2.4 in  \cite{AK2015}, and consists of the following set of reactions 
\begin{align}\label{eq:867986}
\begin{split}
\emptyset &\overset{\kappa_1}{\to} M\\
M &\overset{\kappa_2}{\to}M +P \\
M &\overset{d_M}{\to} \emptyset\\
P &\overset{d_P}\to \emptyset\\
2P&\overset{\kappa_3}\to D\\
D & \overset{d_D}\to \emptyset,
\end{split}
\end{align}
with $\kappa_1 = 200, \kappa_2 = 10, d_M = 25, d_P = 1, \kappa_3 = 0.01,$ and $d_D= 1$.  
We will assume that each cell/compartment has one gene, and so we suppress that species.  Hence, this is a three species model with $\mathcal{S} = \{M,P,D\}$.  For the compartment model, we have
\begin{align*}
\emptyset   \overset{\kappa_I}{\underset{\kappa_E}{\rightleftarrows}} C \overset{\kappa_F}{\underset{\kappa_C}{\rightleftarrows}} 2C,
\end{align*}
 with $\kappa_I = 10$, $\kappa_E = 3.5$,  $\kappa_F = 3$, and $\kappa_C = 0.1$.  
 We will initialize the RNIC model with a single compartment with zero copies of each species.  Moreover, we will initialize each compartment that appears with a reaction network with zero copies of each species.  Finally, we will use our methods to estimate 
\[
\frac{d}{d\kappa_2} \E[ \text{Total dimers}]\quad \text{ and } \quad \frac{d}{d\kappa_F} \E[ \text{Total dimers}], 
\]
where ``Total dimers'' is the sum of the number of dimers across all compartments.

Unlike the previous model, this RNIC model is not amenable to analysis.  For the parameters chosen, the compartment model has a limiting mean of approximately 7.4 compartments (plots not shown).  However, the fluctuations around this mean can be pronounced.  In Figure \ref{fig:compartmentsGene}, we provide a single realization of the compartment process.

\vspace{6pt}

\noindent \textbf{Estimation of $\frac{d}{d\kappa_2} \E[\text{Total dimers}]$.} 
Because the process is not amenable to analysis (which, of course, is the standard situation when one is using simulation and Monte Carlo), we must choose an $h$ for our finite difference methods that we believe limits our bias sufficiently.  For the parameter $\kappa_2$, we chose to use a centered finite difference with $h = 1/10$, which is 1\% of the nominal value of 10.  We chose this by observing the behavior of the estimators for $h = 1$ and $h = 1/10$ using centered finite differences for the Split Coupling with $n =$ 10,000 paths and not seeing any difference in the estimated values for the sensitivity (plots not shown).  We then chose $h = 1/10$ to be conservative.  Note that such a trial and error approach is often  necessary when using finite difference methods as one does not \textit{a priori} know the bias. This extra computational cost should be taken into account when deciding between finite difference methods and unbiased methods such as the Girsanov method.

\begin{figure}
\begin{center}
\includegraphics[width=3.0in]{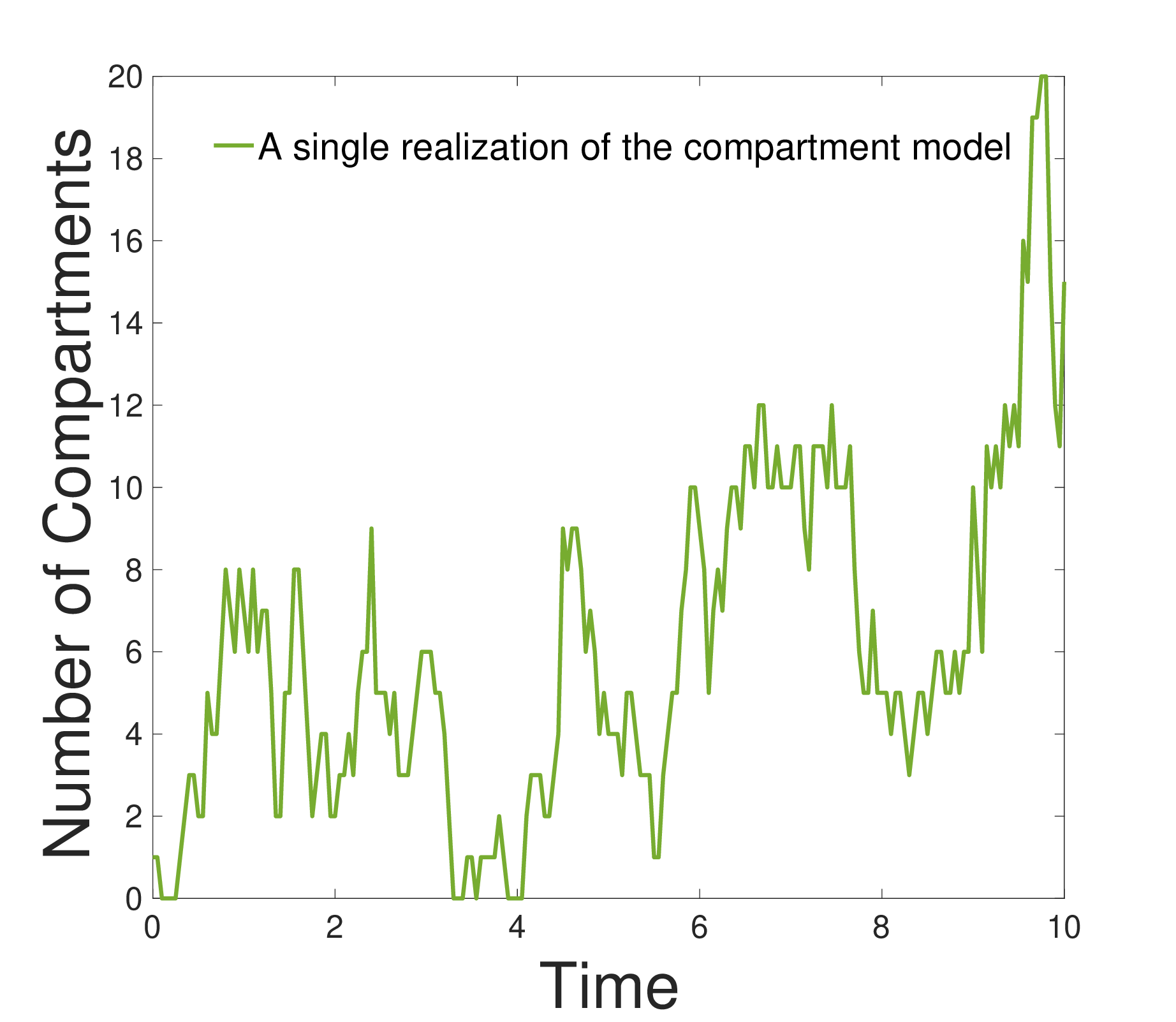}
\end{center}
\caption{A  realization of the compartment model for the birth and death process of Example \ref{example:gene}.}
\label{fig:compartmentsGene}
\end{figure}

See Figure \ref{fig:2345670} for the variances of the different methods when estimating $\frac{d}{d\kappa_2} \mathbb{E}[\text{Total dimers}]$. 
\begin{figure}
\begin{center}
\includegraphics[width=4.5in]{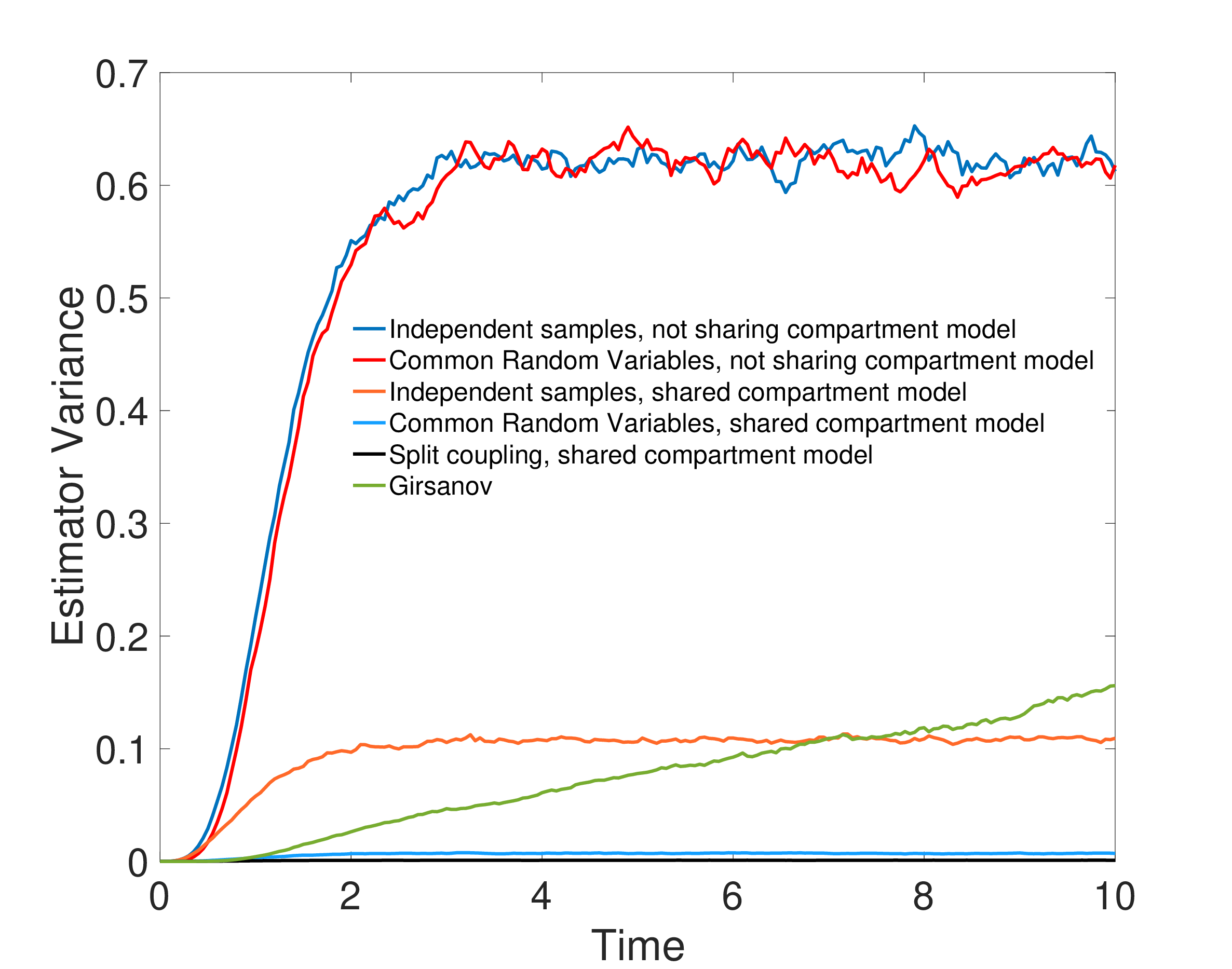}
\end{center}
\caption{Estimator variances \eqref{eq:estimatorvarianceformula} for  $\frac{d}{d\kappa_2} \E[ \text{Total dimers}]$ using various methods.  For the (centered) finite difference methods, we used a perturbation of $h = 1/10$ and $n =$ 10,000 coupled paths.  For the Girsanov method we used $n=$ 20,000 paths.   We leave out the local-CRP method since it is indistinguishable from the Split Coupling and the $x$-axis.}
\label{fig:2345670}
\end{figure}
We note that, as expected, the simple act of sharing the compartment model provides the lion share of the variance reduction.  We also note that the Split Coupling provides an estimator with such a low variance that it is difficult to differentiate from the $x$-axis.  We therefore zoom in and provide in Figure \ref{fig:5yu8976} a plot of the three  methods with the lowest estimator variance.  In each, the compartment model is shared between the coupled processes (as detailed in Algorithm \ref{alg:coupling_RN}).  The chemical models are coupled in the following ways: Common Random Variables (CRV), local Common Reaction Path (local-CRP), and the Split Coupling. As predicted by the work in \cite{anderson2015asymptotic}, and as seen in the previous example, the local-CRP method and the Split Coupling method provide nearly identical variances.  
Finally, in Table \ref{table:56467988}, we provide the simulation times and random variables  required for the various methods. As in the previous example, the Split Coupling, with shared compartments, is  the most efficient.  The relative slowness of the local-CRP method could possibly be alleviated via a different implementation in a different programing language, but it will presumably never be more efficient than the split coupling for this particular problem.

\begin{figure}
\begin{center}
\includegraphics[width=4.5in]{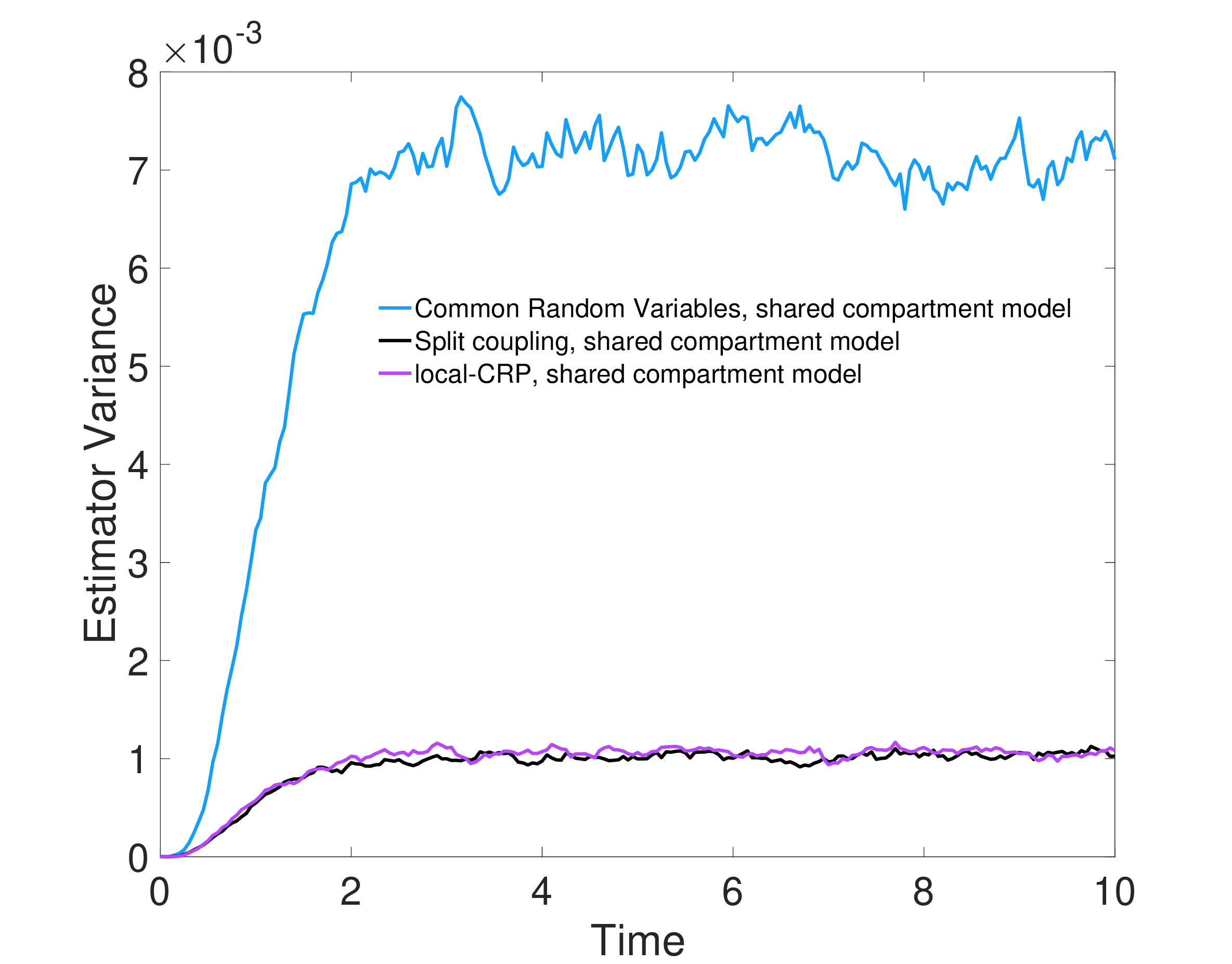}
\end{center}
\caption{Estimator variance \eqref{eq:estimatorvarianceformula} for  $\frac{d}{d\kappa_2} \E[ \text{Total dimers}]$ using various methods.  Each method shares the compartment model.  However, the chemical models are coupled in various ways including:   Common Random Variables, local-CRP, and the Split Coupling.  Note that this figure simply zooms in on the plots for  Common Random Variables and the Split Coupling  from Figure \ref{fig:2345670}, while adding the plot for the local-CRP method.}
\label{fig:5yu8976}
\end{figure}

\begin{table}
    \centering
    \begin{tabular}{|l|c|c|}\hline
        \hspace{2in} Method & Time  & RVs used\\
        \hline
        Girsanov & 308 seconds & $1.5\times 10^9$ \\\hline
        Independent samples, not sharing compartment model & 302 seconds & $1.50\times 10^9$ \\\hline
        Common Random Variables, not sharing compartment model & 322 seconds &  $1.50\times 10^9$ \\\hline
        Independent samples, shared compartment model  & 300 seconds & $1.47\times 10^9$ \\\hline
        Common Random Variables, shared compartment model  & 368 seconds & $1.48\times 10^9$ \\\hline
        local-CRP,  shared compartment model & 697 seconds & $1.54 \times 10^9$  \\\hline
        Split Coupling,  shared compartment model &  298 seconds & $7.57\times 10^8$ \\\hline
    \end{tabular}
    \caption{Simulation times for the various methods for the estimation of $\frac{d}{d\kappa_2} \E[\text{Total dimers}]$.    $n=$ 10,000 coupled paths were used for the finite difference methods and $n=$ 20,000 paths were used for the Girsanov method.}
\label{table:56467988}
\end{table}

\vspace{6pt}

\noindent \textbf{Estimation of $\frac{d}{d\kappa_F} \E[ \text{Total dimers}]$.} 
After using the same selection process for the perturbation, we once again chose a perturbation that is 10\% of the nominal value: $h = 3/10$.  We also again simulated $n=$ 10,000 paths of the coupled processes and 20,000 paths for the Girsanov method.   
Figure \ref{fig:87978675} provides plots of the estimator variance for each method.    The simulation times and numbers of random variables used are similar to those found in Table \ref{table:56467988}, so we omit them.
We note that up through time 2, the Girsanov method has a lower variance than the split coupling method.  However, the variance  of the Girsanov method increases monotonically as time increases and quickly becomes significantly higher than the Split Coupling.

\begin{figure}
\begin{center}
\includegraphics[width=4.5in]{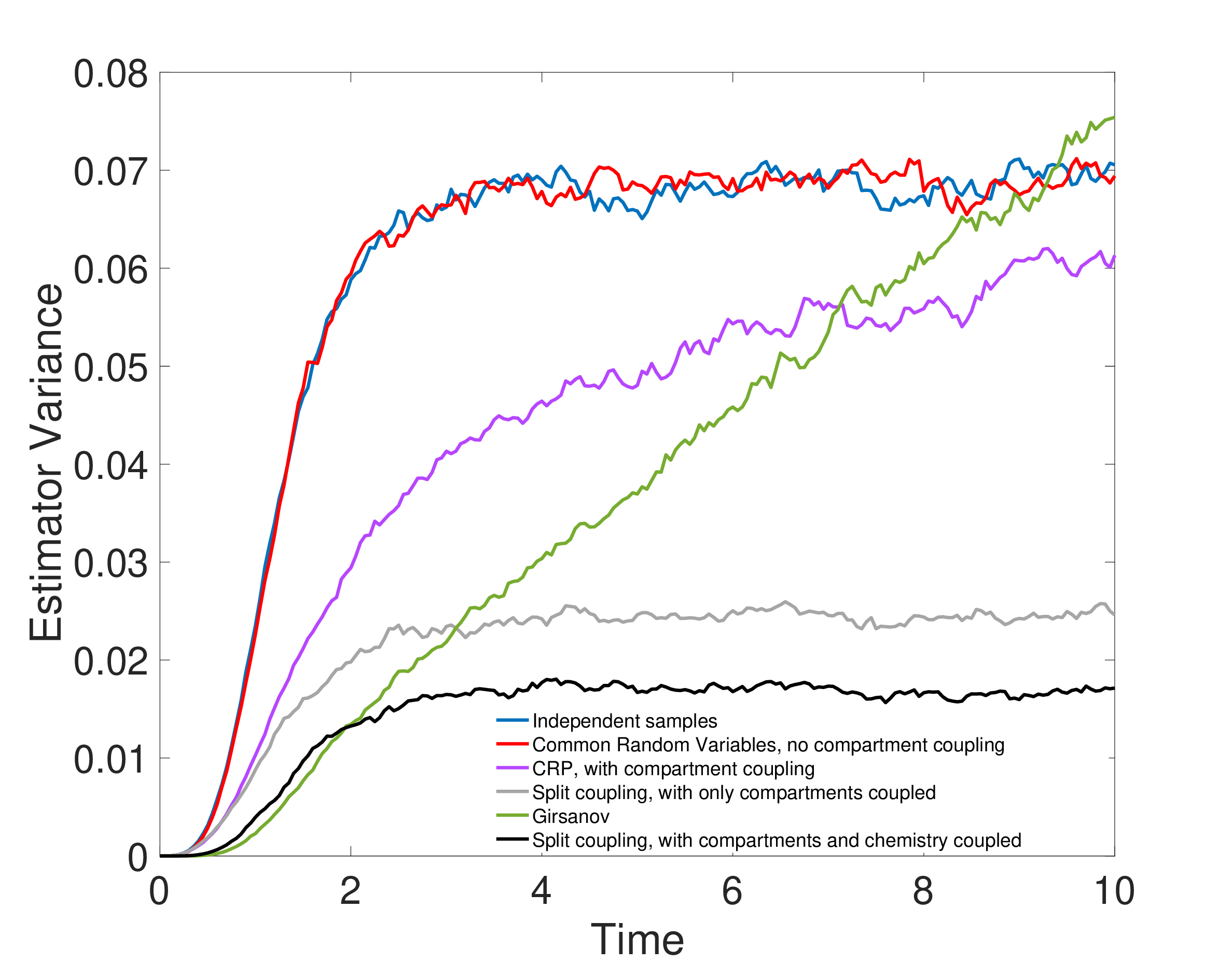}
\end{center}
\caption{
Estimator variance \eqref{eq:estimatorvarianceformula} for  $\frac{d}{d\kappa_F} \E[ \text{Total dimers}]$ using various methods.  For the (centered) finite difference methods, we used a perturbation of $h = 3/10$ and $n =$ 10,000 paths.  For the Girsanov method we used $n=$ 20,000 paths.}
\label{fig:87978675}
\end{figure}


We  also note that the classical CRP method provides a variance that appears to be growing.  This behavior is in line with the observations made in \cite{AndCFD2012}, where it was observed that the processes generated via the classical CRP method decouple as time increases, in which case the variance of the classical CRP method should approach that of independent samples.  

\vspace{6pt}

\noindent \textbf{The many parameter setting.}  
Thus far, we have assumed that $\theta$ is one-dimensional. Consider now the case that $\theta \in \mathbb{R}^K$.  In this case, our implementations would call for $K+1$ path generations when we use a CRP method (either the local or classical), including the nominal process and each of the perturbed processes.  However, the Split Coupling would require the generation of $2K$ paths (since we need a nominal process for each perturbation).  If $K$ is large, the CRP methods will overtake the Split Coupling in terms of efficiency.

To explore this, we consider again the mRNA-protein-dimer model, except now simultaneously perturb all $K=6$ of the chemistry parameters.  Because we are only perturbing the chemical parameters, we utilize the local-CRP coupling.  We still use $n=$ 10,000 paths for each method and simulate to a final time of 10.  Also, for each parameter, we perturb by 10\% of the nominal value (this choice does not affect simulation times or numbers of random variables used, which is our interest here).  Our finding are summarized below.
\begin{itemize}
\item The local-CRP method took 4,401 seconds and required $5.31 \times 10^9$ random variables.  These numbers are 6.3 and 3.45 times the values found when only a single parameter is perturbed (as reported in Table \ref{table:56467988}).  Note that $\frac{6+1}{2} = 3.5$ was the crude estimate for the  factor of the increase in computational complexity.

\item The Split Coupling method took 2,635 seconds and required $4.52 \times 10^9$ random variables.  These numbers are 8.8 and 5.97 times the values found when only a single parameter is perturbed (as reported in Table \ref{table:56467988}).  Note that $\frac{2*6}{2} = 6$ was the crude estimate for the factor of the increase in computational complexity.
\end{itemize}
Taken together, we see that as the number of parameters being perturbed increases, the CRP method does, in fact, gain in efficiency as compared to the Split Coupling.  When $K$ is extremely large (with the specifics depending upon the problem at hand) the CRP methods, and the local-CRP method in particular, will become the most efficient.  \hfill $\square$
\end{example}

\section{Discussion}
\label{sec:discussion}

In this paper, we have provided numerous methods utilizing Monte Carlo  for parametric sensitivity analysis for RNIC models, as introduced in \cite{DZ,anderson2023stochastic}.  In particular, we have synthesized 
work related to finite difference and Girsanov transform methods for parameteric sensitivity analysis for stochastic reaction networks and succinctly developed each of the relevant methods for this new, and more complex, modeling choice.  

In terms of guidance, it seems clear that when only a small number of parameters is being perturbed, the Split Coupling is the most efficient studied here.  However, as the number of parameters being perturbed increases, the CRP methods (and especially the local-CRP in cases when the parameters are from the chemical model) becomes the most efficient.  When the shift happens (between the Split Coupling and the CRP methods) will be problem specific.

There are multiple avenues for future work. First, the RNIC model introduced here is the most ``basic'' version of the model, however it can be generalized in a number of ways.  For example, it is possible to allow the transition rates of the compartment model (e.g., the fragmentation rate) to depend upon the contents of the compartments \cite{DZ,AHR2024}.  It is not a priori obvious how to extend each of the developed methods to this different model in the most efficient manner possible.  However, we hope that the extensions provided here, for the basic RNIC model, serve as a strong starting point.

Second, there are myriad other methods that have been developed for the estimation of parametric sensitivities for stochastic reaction networks that have not been extended to the RNIC framework within this paper.  For example, we have not attempted to extend the 
 Auxiliary Path Algorithm, \cite{gupta2013unbiased} the Poisson Path Algorithm, \cite{gupta2014efficient} the hybrid pathwise approach,\cite{wolf2015hybrid} the  Integral Path Algorithm \cite{gupta2018estimation}, or the centered Girsanov transformation method\cite{warren2012steady}, each of which is unbiased.  We feel that each of these methods falls outside  the scope of the current work, which largely focused on the development of coupling methods.  
  It is possible that extensions of any of these methods could prove beneficial.
  
Third, we have not attempted to extend methods, such as that found in \cite{AndSkubak}, for the estimation of second derivatives.  Extending such methods is certainly worthy of consideration.

Finally, there is a significant amount of analysis that could be carried out to better understand the finite difference methods introduced here.  Unbiased Monte Carlo estimators, such as the Girsanov method, will converge to the true value of the quantity being estimated, in the sense of confidence intervals (or $\mathcal L_2$ error--i.e., accounting for both bias and estimator standard deviation) at a rate of  $O(n^{-1/2})$, as $n \to \infty$, where $n$ is the number of sample paths.  
For the centered finite difference methods, which have a bias that must be taken into account, determining the $\mathcal{L}_2$ convergence rate is more difficult.  In the setting of reactions networks, the convergence rates can be $O(n^{-1/3})$ or $O(n^{-2/5})$, or something else, depending upon how $Var(f(X^{\theta+h/2}(t)) - f(X^{\theta-h/2}(t)))$ scales, as $h\to 0$. See section 2.1 of Ref. 26 for a detailed discussion of this topic.
Determining how $Var(f(X^{\theta+h/2}(t))-f(X^{\theta-h/2}(t)))$ scales, as $h \to 0$, for different finite difference methods in the case where $X$ is an RNIC model would be a valuable piece of future work that could more rigorously determine when various methods should be used.  Additionally, it is important to investigate how this scaling depends on the properties of the function $f$ and the specific characteristics of the process $X$.

\vspace{.1in}

\noindent \textbf{Acknowledgements}

Anderson gratefully acknowledges support from NSF grant DMS-2051498.
Howells gratefully acknowledges support from MUR PRIN grant number 2022XRWY7W.

\appendix

\section{Monte Carlo and finite difference methods for parametric sensitivities}
\label{sec:MonteCarlo}

The basic theory of Monte Carlo and finite difference methods for parametric sensitivities can be found in various papers and textbooks.  No originality for this material is claimed.

Let $X^\theta: \mathbb{R}_{\ge 0} \to \mathbb{X}$ be a family, parameterized by $\theta \in \RR$, of continuous-time stochastic processes with state space $\mathbb{X}$.  In the setting of the current paper, $X^\theta$ is a continuous-time Markov chain with a discrete state space.  We are assuming $\theta$ is one-dimensional here, but everything extends in the obvious manner if $\theta \in \RR^K$ for some finite, positive integer $K$.

Let $f : \mathbb{X} \to \RR$ be a function of the state of the system that gives a measurement of interest and define 
 \begin{equation*}
   J(\theta) := \E\left[ f(X^{\theta}(t))\right].
\end{equation*}
 The problem of interest is to efficiently estimate $\frac{d}{d \theta}J(\theta) = J'(\theta)$, and to do so using finite difference methods with Monte Carlo.  (The other method utilized in this paper, a Girsanov or Likelihood Transformation method, is discussed in Section \ref{sec:girsanov}.)
 
 In order to estimate $J'(\theta)$, it is natural to utilize a centered finite difference:  
\begin{equation}
	J'(\theta) \approx \frac{\E\left[ f(X^{\theta + h/2}(t))\right] - \E \left[f(X^{\theta - h/2}(t))\right]}{h},
	\label{eq:centered}
\end{equation}
as its bias is $O(h^2)$, as $h\to 0$, see  \cite{GlynnAsmussen2007}. 
This should be compared with the forward difference, which has a bias of $O(h)$
\begin{align*}
	J'(\theta) &= \frac{\E\left[f(X^{\theta + h}(t)\right]) - \E \left[ f(X^{\theta}(t))\right] }{h} + O(h).
\end{align*}
The estimator for \eqref{eq:centered} using centered finite differences is
\begin{equation}
D_n(h) = \frac{1}{n} \sum_{\ell = 1}^n d_{[\ell]}(h), 
\label{eq:difference}
\end{equation}
with
\begin{equation}
	\label{eq:d}
	d_{[\ell]}(h)= \frac{f(X_{[\ell]}^{\theta + h/2}(t)) - f(X_{[\ell]}^{\theta - h/2}(t))}{h},
\end{equation}
where $X_{[\ell]}^{\theta}$ represents the $\ell$th path generated with parameter choice $\theta$,  $n$ is the number of paths generated, and the $d_{[\ell]}(h)$ are generated independently.

Many computations are performed with a target variance (which yields a target size of the confidence interval).  Denoting the target variance by $V^*$, we see that the number of paths required is  approximated by the  solution to
\[
	\mathsf{Var}\left( \frac{1}{n} \sum_{\ell = 1}^n d_{[\ell]}(h) \right) = \frac{1}n \mathsf{Var}(d(h)) = V^* \implies n = \frac{1}{V^*}\mathsf{Var}(d(h)).
\]
Thus,   decreasing the variance of $d(h)$  lowers the computational complexity (total number of computations)  required to solve the problem.

The basic idea of coupling is to lower the variance of $d(h)$ by simulating $X^{\theta + h/2}$ and $X^{\theta - h/2}$ simultaneously, i.e., generating them on the same probability space, so that the two processes are highly correlated or ``coupled.''  That is, instead of generating paths independently, we want to generate a pair of paths $(X^{\theta + h/2}, X^{\theta - h/2})$ so that the variance of $f(X^{\theta + h/2}(t)) - f(X^{\theta - h/2}(t))$ is reduced.  The basic idea of any such coupling is to reuse, or share, some portion of randomness in the generation of each process.   The couplings utilized in this paper, found in Section \ref{sec:methods}, include using Common Random Variables (i.e., simply reusing the seed of the random number generator before generating each path), versions of the Common Reaction Path method \cite{Khammash2010} and local-Common Reaction Path method,\cite{anderson2015asymptotic} and a version of the Split Coupling. \cite{AndCFD2012}

That all said, it is not enough to simply provide a low variance estimator without taking the bias of the method into account.  Thinking in terms of confidence intervals, it is clear that there is little reason to desire a standard deviation for an estimator that is significantly less than  its bias.  This  idea can be made precise by considering the $\mathcal{L}_2$ error of an estimator.  For example, if we denote the bias of our estimator \eqref{eq:difference} by $B(h) = |J'(\theta) - \E[D_n(h)]|$, then the square of the $\mathcal{L}_2$ error is
\begin{align*}
 \E\left[ | J'(\theta) - D_n(h)|^2\right] &= \E\left[ | J'(\theta) - \E[D_n(h)] + \E[D_n(h)]  - D_n(h)|^2\right]= B(h)^2 + \text{Var}(D_n(h)).
\end{align*}
  Moreover, as noted above, as $h \to 0$ the bias shrinks for our finite difference methods but the variance tends to infinity (because of the division by $h$).  Hence, there is a subtle bias-variance tradeoff that must be accounted for with these methods.

\section{Standard algorithms for stochastic reaction networks}
\label{sec:MC_Algorithms}

We provide the Gillespie algorithm \cite{Gill76,Gill77} and the next reaction method \cite{Gibson2000, Anderson2007a}, which are the two exact simulation methods that are most widely used.  We also provide a Gillespie version of the Girsanov transformation method.

We remind that throughout this work, we  assume a reaction network with reaction vectors denoted by $\zeta_k \in \ZZ^d$ and intensity functions (or rate functions) denoted by $\lambda_k$, where we have indexed the reactions by $k$.  For the sake of clarity, we will denote (in this section only) the number of reaction types by $R$.

We first provide Gillespie's algorithm \cite{Gill76,Gill77}.
\begin{algorithm}[Gillespie's algorithm]
\label{alg:Gillespie}
Given: a reaction network with reaction vectors $\{\zeta_k\}_{k=1}^R$ and intensity functions $\{\lambda_k\}_{k=1}^R$, an initial distribution, $\mu$, on $\ZZ^d_{\ge 0}$.

Repeat steps 3 -- 8 until a stopping criteria is reached. All calls to random variables are independent from all others.
\begin{enumerate}[itemsep=-.05ex]
\item Determine $X(0)$ via $\mu$.
\item Set $t = 0$.
\item For each $k\in \{1,\dots,R\}$, determine $\lambda_k(X(t))$ and set $\lambda_{\text{tot}} = \sum_{k=1}^R \lambda_k(X(t))$.
\item Let $u_1,u_2$ be independent random variables that are uniformly distributed on $[0,1]$.
\item Set $\Delta = -\ln(u_1)/\lambda_{\text{tot}}$.
\item Find $\ell \in \{1,\dots, R\}$ so that
\[
	\frac{1}{\lambda_{\text{tot}}}\sum_{k = 1}^{\ell -1} \lambda_k(X(t)) \le u_2 < \frac{1}{\lambda_{\text{tot}}}\sum_{k = 1}^{\ell } \lambda_k(X(t)).
\]
\item Set $X(t+\Delta) = X(t) + \zeta_\ell$.
\item Set $t \leftarrow t + \Delta$.
\end{enumerate}
\end{algorithm}

We now provide the next reaction method, as it appeared in \cite{Anderson2007a}.  Note that this algorithm is simply a method for simulating  Kurtz's representation \eqref{eq:RTC}.  Essentially the same algorithm (though with the addition of the usage of index priority queues)  appeared  in \cite{Gibson2000}.

\begin{algorithm}[Next Reaction method \cite{Anderson2007a}]
\label{alg:NRM}

Given: a reaction network with reaction vectors $\{\zeta_k\}_{k=1}^R$ and intensity functions $\{\lambda_k\}_{k=1}^R$, an initial distribution, $\mu$, on $\ZZ^d_{\ge 0}$.

Repeat steps 5 -- 10 until a stopping criteria is reached. All calls to random variables are independent from all others.

\begin{enumerate}[itemsep=-.05ex]
\item Determine $X(0)$ via $\mu$.
\item Set $t = 0$.
\item For each $k\in \{1,\dots, R\}$, set $T_k = 0$.
\item Let $\{e_k\}_{k = 1}^R$ be a collection of independent unit-exponential random variables and for each $k\in \{1,\dots, R\}$, set $P_k = e_k$.
\item For each $k\in \{1,\dots,R\}$, determine $\lambda_k(X(t))$.
\item Find the minimum of the values $\left\{ \frac{P_k - T_k}{\lambda_k(X(t))}\right\}_{k=1}^R$.  Denote the minimum by $\Delta$ and denote the index of the minimum value by $\ell$.
\item Set $X(t+\Delta) = X(t) + \zeta_\ell$.
\item For each $k\in \{1,\dots,R\}$, set $T_k = \lambda_k(X(t)) \cdot \Delta$.
\item Set $P_\ell \leftarrow P_\ell + e_0$, where $e_0$ is a unit-exponential random variable (independent from previous).
\item Set $t\leftarrow t +\Delta$.

\end{enumerate}
\end{algorithm}

We turn to the Girsanov transformation method, often called the Likelihood Transformation method outside of the biosciences.  See either \cite{Plyasunov2007} or \cite{GlynnAsmussen2007} for relevant references. 
For concreteness, we will assume stochastic mass-action kinetics and we suppose that we are computing the derivative of $\E[f(X(t_{\text{end}}))]$ with respect to the rate parameter $\kappa_j$, where $t_{\text{end}}$ is some fixed time.
We will denote the required weight function by $W(t)$.
Therefore, the output of the algorithm consists of both $X(t_{\text{end}})$ and $W(t_{\text{end}})$, which can be used as an unbiased estimator $\frac{\partial}{\partial \kappa_j} \E[f(X(t_{\text{end}}))] \approx \frac{1}{n}\sum_{\ell=1}^n f(X_{[\ell]}(t_{\text{end}}))\cdot W_{[\ell]}(t_{\text{end}})$, where the subscript $[\ell]$ enumerates the independent realizations.

\begin{algorithm}[Girsanov transformation method]
\label{alg:GirsanovStandard}
Given: a reaction network with reaction vectors $\{\zeta_k\}_{k=1}^R$ and intensity functions $\{\lambda_k\}_{k=1}^R$, an initial distribution, $\mu$, on $\ZZ^d_{\ge 0}$.

All calls to random variables are independent from all others.
\begin{enumerate}[itemsep=-.05ex]
\item Determine $X(0)$ via $\mu$.
\item Set $t = 0$ and $W(0) = 0$.
\item For each $k\in \{1,\dots,R\}$, determine $\lambda_k(X(t))$ and set $\lambda_{\text{tot}} = \sum_{k=1}^R \lambda_k(X(t))$.
\item Let $u_1,u_2$ be independent random variables that are uniformly distributed on $[0,1]$.
\item Set $\Delta = -\ln(u_1)/\lambda_{\text{tot}}$.
\item If $t + \Delta > t_{\text{end}}$, do the following (otherwise, proceed to step 7):
\begin{enumerate}
	\item set $W(t_{\text{end}}) = W(t) - (t_{\text{end}} - t)\cdot \frac{\lambda_j(X(t))}{\kappa_j},$
	\item set $X(t_{\text{end}}) = X(t)$,
	\item  break from the algorithm and report $X(t_{\text{end}})$ and $W(t_{\text{end}})$.
\end{enumerate}
\item Find $\ell \in \{1,\dots, R\}$ so that
\[
	\frac{1}{\lambda_{\text{tot}}}\sum_{k = 1}^{\ell -1} \lambda_k(X(t)) \le u_2 < \frac{1}{\lambda_{\text{tot}}}\sum_{k = 1}^{\ell} \lambda_k(X(t)).
\]
\item If $\ell = j$, set 
\[
W(t+\Delta) = W(t) + \frac{1}{\kappa_j} -  \frac{\lambda_j(X(t))}{\kappa_j} \cdot \Delta
\]
otherwise, if $\ell \ne j$, set
\[
	W(t+ \Delta) = W(t) -  \frac{\lambda_j(X(t))}{\kappa_j} \cdot \Delta.
\]
\item Set $X(t+\Delta) = X(t) + \zeta_\ell$.
\item Set $t \leftarrow t + \Delta$, and return to step 3.
\end{enumerate}
\end{algorithm}

 \bibliographystyle{plain}
 \bibliography{RNIC_COUPLING}

\end{document}